\documentclass[aps,prd,twocolumn,preprintnumbers,showpacs,nofootinbib,amssymb]{
revtex4}
\usepackage{graphicx}
\usepackage{amsmath}
\usepackage{amssymb}
\usepackage{amsfonts}
\usepackage{bm}
\usepackage{color}

\def\be{\begin{equation}}
\def\ee{\end{equation}}
\def\bea{\begin{eqnarray}}
\def\eea{\end{eqnarray}}

\begin{document}

\title{Quantum tunneling rate of dilute axion stars close to the maximum mass}
\author{Pierre-Henri Chavanis}
\affiliation{Laboratoire de Physique Th\'eorique, Universit\'e de Toulouse,
CNRS, UPS, France}

\begin{abstract}

We compute the quantum tunneling rate of dilute
axion stars close to the maximum mass [P.H. Chavanis, Phys.
Rev. D {\bf 84}, 043531 (2011)] using the theory of instantons. We
confirm that the lifetime of metastable states is extremely long, scaling as
$t_{\rm life}\sim e^N\, t_D$ (except close to the critical point),
where $N$ is
the number of axions in the system and $t_D$ is the
dynamical time ($N\sim 10^{57}$ and $t_D\sim 10\, {\rm hrs}$
for
typical QCD axion stars; $N\sim 10^{96}$ and $t_D\sim 100\, {\rm
Myrs}$ for the quantum core of a dark matter halo made of ultralight axions).
Therefore,
metastable equilibrium states can be considered as stable equilibrium states in
practice. We develop a finite size scaling theory close to the maximum mass and
predict that the collapse time at criticality scales as $t_{\rm
coll}\sim N^{1/5}t_D$ instead of being infinite as when fluctuations are
neglected. The collapse time
is smaller than the age of the universe for QCD axion stars and larger than the
age of the universe for dark matter cores made of ultralight axions. We also
consider the thermal tunneling
rate and reach the same conclusions. We
compare our
results with similar results obtained for Bose-Einstein
condensates in laboratory, globular clusters in astrophysics, and quantum field
theory in the early Universe. 
\end{abstract}


\maketitle

\section{Introduction}
\label{sec_introduction}

The nature of dark matter (DM) is still unknown and constitutes
one of the
greatest
mysteries of modern cosmology. The  cold dark
matter (CDM) model in which DM is assumed to be made of weakly
interacting massive particles (WIMPs) of mass
$m\sim {\rm
GeV/c^2}$  works
remarkably well at large
(cosmological) scales \cite{ratra} but encounters problems at small
(galactic) scales.
These problems are known as the cusp problem \cite{cusp}, the missing
satellite
problem \cite{satellites}, and the too big to fail problem \cite{tbtf}. In
addition, there is no current evidence for any CDM particle such as the WIMP. In
order to
solve this ``CDM crisis'', it has been proposed to take the quantum nature of
the
particles into account. For example, it has been suggested that DM may be
made of bosons in the form of Bose-Einstein condensates (BECs) at
absolute zero
temperature
\cite{baldeschi,khlopov,membrado,bianchi,sin,jisin,leekoh,schunckpreprint,
matosguzman,sahni,
guzmanmatos,hu,peebles,goodman,mu,arbey1,silverman1,matosall,silverman,
lesgourgues,arbey,fm1,bohmer,fm2,bmn,fm3,sikivie,mvm,lee09,ch1,lee,prd1,prd2,
prd3,briscese,
harkocosmo,harko,abrilMNRAS,aacosmo,velten,pires,park,rmbec,rindler,lora2,
abrilJCAP,mhh,lensing,glgr1,ch2,ch3,shapiro,bettoni,lora,mlbec,madarassy,marsh,
abrilph,playa,stiff,souza,freitas,alexandre,schroven,pop,cembranos,
schwabe,fan,calabrese,bectcoll,chavmatos,hui,abrilphas,
chavtotal,shapironew,mocz,zhang,suarezchavanisprd3,veltmaat,moczSV,phi6,bbbs,
cmnjv,psgkk,ekhe,matosbh,tkachevprl2,epjpbh,ag,lhb,nmibv,zlc,bm,modeldm,bft,
bblp,dn,bbes,bvc,moczamin,mabc,gga,moczprl,mcmh,mcmhbh,reig,dm,adgltt,
moczmnras,verma,braxbh,lancaster}
(see the
Introduction of \cite{prd1} and Ref. \cite{leerevue} for an early history of
this model and Refs. \cite{srm,rds,chavanisbook,marshrevue,niemeyer}
for reviews). In this
model, DM halos are
interpreted as gigantic boson stars described by a scalar field (SF) that
represents the wavefunction $\psi$ of the BEC. The mass of the DM boson has to
be very small (see below) for quantum mechanics to manifest itself at galactic
scales. By contrast, quantum mechanics is completely negligible at astrophysical
scales for ``heavy'' particles of mass $m\sim {\rm GeV/c^2}$ such as WIMPs.

One possible DM particle candidate is the axion
\cite{kc}.  Axions are hypothetical
pseudo-Nambu-Goldstone bosons of the
Peccei-Quinn \cite{pq} phase transition associated with a $U(1)$ symmetry that
solves the strong charge parity (CP) problem of quantum chromodynamics (QCD).
The QCD axion is a spin-$0$ particle with a very
small
mass  $m=10^{-4}\,
{\rm eV}/c^2$ and an extremely weak self-interaction $a_s=-5.8\times
10^{-53}\, {\rm m}$  arising
from
nonperturbative effects in QCD ($a_s$ is the scattering length of the axion)
\cite{weinbergaxion,wilczekaxion}. 
Their role in cosmology has
been first
investigated in \cite{preskill,abbott,dine,davis}. 
 Axions have
huge occupation numbers so they can be described by a classical 
relativistic quantum field theory with a real scalar field $\varphi({\bf r},t)$
whose
evolution is governed by the Klein-Gordon-Einstein (KGE) equations. In the
relativistic regime, the particle number is not conserved. In the
nonrelativistic
limit, axions can
be described by an effective field theory with a complex scalar field $\psi({\bf
r},t)$
whose evolution is governed  by the Gross-Pitaevskii-Poisson (GPP) equations
(see Appendix \ref{sec_kge}). In the
nonrelativistic regime, the particle number is conserved.
One
particularity of the QCD axion is to have a negative scattering length ($a_s<0$)
corresponding
to an attractive self-interaction.

The formation of structures in an axion-dominated Universe was first
investigated by
Hogan and Rees \cite{hr} and Kolb and Tkachev \cite{kt}.
In
the very early
Universe, the axions are relativistic but self-gravity can be neglected with
respect to their attractive self-interaction. These authors found that the
attractive self-interaction of the axions generates  very dense structures
corresponding to pseudo-soliton configurations  that they called ``axion
miniclusters'' \cite{hr} or ``axitons'' \cite{kt} (these
nongravitational solitons are also called ``oscillons''). These
axitons
have a mass
$M_{\rm
axiton}\sim 10^{-12}\, M_{\odot}$ and a radius $R_{\rm
axiton}\sim 10^{9}\, {\rm m}$. 
At later times, self-gravity must be taken into account. Kolb and Tkachev
\cite{kt} mentioned the possibility to form
boson stars\footnote{Boson stars, that are the solutions of the 
KGE equations, were introduced by Kaup \cite{kaup} and
Ruffini and Bonazzola \cite{rb} in the case where the bosons have no
self-interaction. Boson stars in which the bosons have a repulsive
self-interaction ($a_s>0$) were considered
later by Colpi {\it et al.} \cite{colpi} using field theory and by Chavanis and
Harko \cite{chavharko} using a hydrodynamic treatment valid in the
Thomas-Fermi (TF) limit. These
authors showed that boson stars can exist only below a maximum mass, $M_{\rm
max}=0.633\,{\hbar c}/{Gm}$ for noninteracting bosons and $M_{\rm max}=0.307
\,\left
({a_s\hbar^2 c^4}/{G^3m^3}\right )^{1/2}$ for bosons with a repulsive
interaction in the TF limit, due to
general relativistic effects.} by Jeans instability.
This possibility was originally proposed by Tkachev \cite{tkachev,tkachevrt} who
introduced the names ``gravitationally bound axion condensates''
\cite{tkachev} and ``axionic Bose stars'' \cite{tkachevrt}, becoming later
``axion stars''. Tkachev \cite{tkachev,tkachevrt} and Kolb and Tkachev \cite{kt}
discussed the maximum mass of these axion stars due to general relativity
but, surprisingly, they considered the case of a repulsive self-interaction
($a_s>0$). Since axions have an
attractive self-interaction ($a_s<0$), their result does not apply to
axion stars.

The case of boson stars with an attractive
self-interaction ($a_s<0$), possibly representing axion stars, has been
considered only
recently
\cite{prd1,prd2,bectcoll,phi6,epjpbh,mcmh,mcmhbh,gul0,gu,bb,ebyinfrared,guth,
ebybosonstars,braaten,braatenEFT,davidson,ebycollapse,bbb,ebylifetime,cotner,
ebycollisions,ebychiral,tkachevprl,helfer,svw,visinelli,moss,ebyexpansion,ebybh,
ebydecay,namjoo,ebyapprox,nsh,nhs,chs,croon,ebyclass,elssw,guerra} (see a
review in \cite{braatenrevue}). The Jeans instability of
a Newtonian self-gravitating BEC with an attractive $|\psi|^4$
self-interaction was studied by Chavanis \cite{prd1,aacosmo} and Guth {\it et
al.} \cite{guth}.  An infinite homogeneous BEC of axions is unstable to the
formation of localized denser clumps of axions. The clumps can be axitons
bound by axion self-interaction or axion stars bound by self-gravity. In the
case of
axion stars, gravitational cooling \cite{seidel94,gul0,gu}
provides an efficient mechanism for relaxation to a stable
configuation. The existence of a maximum mass
for axion stars was envisioned by 
Barranco and Bernal \cite{bb} but they did not determine this critical
mass.\footnote{Barranco and Bernal
\cite{bb} developed a general relativistic formalism based on the KGE equations
which is well-suited to the limit where $|a_s|$ is very small (or the axion
decay constant $f$ is close to the Planck energy $M_Pc^2$). However, their
scaling relations were not adapted to study dilute axion stars for which
$f\ll M_Pc^2$ and they could not explore this type
of stars thoroughly, nor determine their maximum mass.
The solutions that they found ($M\sim 10^{14}\, {\rm kg}$ and
$R\sim 10\, {\rm m}$ for QCD axions) have a mass much lower than the maximum
mass $M_{\rm max}=1.29\times 10^{17}\, {\rm kg}$ and correspond
to the
unstable branch $R<R_{99}^*=227\, {\rm km}$ of dilute axion stars (see
below).}
The maximum
mass of
Newtonian self-gravitating BECs with an attractive $|\psi|^4$ self-interaction,
and the
corresponding radius, were first calculated by 
Chavanis and Delfini \cite{prd1,prd2} who obtained the explicit
expressions\footnote{Equivalent expressions, written in terms of
different parameters (e.g. the dimensionless
self-interaction constant $\lambda$ or the axion decay constant $f$), are
given in \cite{phi6}.}
\begin{eqnarray}
\label{intro1a}
M_{\rm max}^{\rm exact}=1.012\, \frac{\hbar}{\sqrt{Gm|a_s|}}
\end{eqnarray}
and
\begin{eqnarray}
\label{intro1b}
(R_{99}^*)^{\rm exact}=5.5\,
\left (\frac{|a_s|\hbar^2}{Gm^3}\right )^{1/2}.
\end{eqnarray}
For $M>M_{\rm max}$ there is no equilibrium
state. For $M<M_{\rm max}$ there
are two possible equilibrium states for the same mass $M$. The solution with
$R>R_{99}^*$ is stable (minimum of energy) while the solution $R<R_{99}^*$ is
unstable (maximum of energy). For $R\gg R_{99}^*$ we are in
the noninteracting limit and for $R\ll R_{99}^*$ we are in
the nongravitational limit. Starting from the KGE equations with the axion
potential, Braaten {\it et al.}
\cite{braaten,braatenEFT} showed
that the results of 
Chavanis and Delfini \cite{prd1,prd2} 
apply to {\it dilute axion
stars} because, for these
objects, it is possible to make the Newtonian approximation and to expand the
axion potential to order $\varphi^4$, leading to the GPP equations with an
attractive $|\psi|^4$ self-interaction (see also Eby
{\it et al.}
\cite{ebyinfrared}, Davidson and  Schwetz
\cite{davidson}, and Appendix
\ref{sec_kge}). Thus,
dilute axion stars can exist only below the maximum
mass $M_{\rm max}$ and above the minimum radius $R_{99}^*$ given by Eqs.
(\ref{intro1a}) and (\ref{intro1b}). We stress that the
maximum mass of dilute axion stars \cite{prd1,prd2} has a nonrelativistic origin
unlike the maximum mass of boson stars \cite{kaup,rb,colpi,chavharko}.

For
QCD axions, the maximum mass $M_{\rm max}^{\rm
exact}=6.46\times 10^{-14}\,
M_{\odot}=1.29\times 10^{17}\, {\rm kg}=2.16\times 10^{-8}\,
M_{\oplus}$ and the corresponding radius $(R_{99}^*)^{\rm exact}=3.26\times
10^{-4}\,
R_{\odot}=227\, {\rm km}=3.56\times 10^{-2}\, R_{\oplus}$ are very small, much
smaller than galactic sizes. Therefore, QCD axions are expected to form mini
axion stars of the size of asteroids (``axteroids'').

However, string theory
\cite{wittenstring}  predicts the existence of axions 
with a very small mass  leading to the notion of string axiverse
\cite{axiverse}. This new class of axions is
called
ultralight axions (ULA) \cite{marshrevue}. 
 For an ULA with a mass 
$m=2.19\times 10^{-22}\, {\rm eV}/c^2$  and a very small attractive
self-interaction $a_s=-1.11\times 10^{-62}\, {\rm fm}$, one finds that the
maximum mass and the minimum radius of 
axionic DM halos are $M_{\rm max}=10^8\, M_{\odot}$ and  $R^*_{99}=1\, {\rm
kpc}$. For smaller (absolute) values of the scattering length, the maximum
mass
is larger. Therefore, ULAs can form giant
BECs with the dimensions of DM halos. These objects may correspond either to
ultracompact DM halos like dwarf spheroidal galaxies (dSphs) or to the
quantum core (soliton) of larger DM halos. In that second case,
the quantum core is surrounded by a halo of scalar radiation (arising from
quantum interferences) resulting from a process of violent
relaxation \cite{lb} and gravitational cooling \cite{seidel94,gul0,gu}. This
``core-halo'' structure has
been evidenced in direct numerical simulations of noninteracting BECDM
\cite{ch2,ch3,moczprl,veltmaat,mocz} and it is
expected to persist for self-interacting bosons.
In the case of ULAs, the quantum core (ground
state of the GPP equations) stems from the
equilibrium
between the quantum pressure (Heisenberg's uncertainty principle), the
attractive self-interaction of the axions and the gravitational attraction. On
the other hand, the ``atmosphere'' has an approximately isothermal \cite{lb} or
Navarro-Frenk-White (NFW) profile \cite{nfw} as obtained in classical numerical
simulations of collisionless matter (see, e.g., \cite{moczSV} for the
Schr\"odinger-Vlasov correspondance). It is the
atmosphere that determines the mass and the size of large DM halos and explains
 why
the halo
radius $r_h$ increases with the halo mass $M_h$ while the core radius
$R_c$ decreases with the core mass $M_c$ (see Appendix L of
\cite{modeldm} for a more detailed discussion). The core mass -- halo
mass relation $M_c(M_h)$ of BECDM halos with an attractive self-interaction has
been determined in \cite{mcmh,mcmhbh}. It is
found that
the core mass
$M_c$ increases with the halo mass $M_h$ up to the maximum mass $(M_c)_{\rm
max}$. Of
course, these core-halo configurations are stable only if
the mass of their core  is smaller than the maximum mass ($M_c<(M_c)_{\rm
max}$). In sufficiently large DM halos, the core mass passes above the maximum
mass, becomes unstable, and undergoes gravitational collapse.

The collapse of dilute axion stars  above $M_{\rm max}$ was first
discussed by Chavanis \cite{bectcoll} using a Gaussian ansatz
and assuming that
the self-interaction is
purely attractive and that the system remains spherically symmetric and
nonrelativistic. In that case, the system is expected to collapse towards a
mathematical singularity (Dirac peak).\footnote{In Ref.
\cite{bectcoll} this mathematical singularity was abusively refered to as a
``black hole''. This terminology is clearly not correct since a
nonrelativistic approach is used in \cite{bectcoll}. What we meant by
``black hole'' was actually a Newtonian ``Dirac peak'' in the sense
of \cite{sp3}.
On the other hand, the Gaussian ansatz used in \cite{bectcoll} provides an
inaccurate description of the late stage of the collapse dynamics. Indeed, in
the
late stage of the collape, the system is dominated by the attractive
self-interaction and the BEC is described by the nongravitational GP equation
with an attractive self-interaction. In that case, it is well-known
\cite{sulem,zakharov} that the collapse is self-similar and leads to a finite
time singularity. The central density becomes infinite in a finite time $t_{\rm
coll}$ at which a singular density profile $\rho\propto r^{-2}$ is formed. The
Dirac peak
may be formed in the post-collapse regime $t>t_{\rm coll}$ as in \cite{sp3}.
This complex
late dynamics cannot be studied with the Gaussian ansatz. However, the Gaussian
ansatz is relevant to determine the collapse time of the system which is
dominated by the early evolution of the system. It is found in 
Ref. \cite{bectcoll} that $t_{\rm coll}\propto (M-M_{\rm max})^{-1/4}$ when
$M\rightarrow M_{\rm max}^+$.} Less idealized scenarios were
considered in later works from numerical simulations. For example, Cotner 
\cite{cotner} showed that the system
may break into several stable pieces (axion ``drops'' \cite{davidson}) of mass
$M'<M_{\rm
max}$, thereby avoiding its catastrophic collapse towards a
singularity. This type of fragmentation has been observed
experimentally in the case of nongravitational BECs with an attractive
self-interaction in a magnetic trap \cite{cornish}. On the other hand,
when the system
becomes dense enough, the   $|\psi|^4$ approximation is not valid anymore and
one
has
to take into account higher order terms in the expansion of the SF potential
(or, better, consider the exact axionic self-interaction potential). These
higher order terms, which can
be repulsive (unlike the $\varphi^4$ term for axions), can account for strong
collisions
between axions. These collisions may have important consequences on the collapse
dynamics. Three possibilities have been considered in the literature:

(i) The first possibility, proposed by Braaten {\it et al.} \cite{braaten}, is
to form a {\it dense axion star} in which the gravitational attraction and the
attractive $\varphi^4$ self-interaction are balanced by the repulsive
$\varphi^6$ (or higher order) self-interaction. They used a
nonrelativistic approximation and determined the mass-radius relation of
axion stars numerically. They recovered the stable branch of dilute axion
stars and the unstable branch of nongravitational axion stars found by
Chavanis and Delfini \cite{prd1,prd2} and evidenced, in addition, a new
stable branch of dense axion stars. On this branch, self-gravity is
negligible (except for very large masses). The mass-radius relation of axion
stars presents therefore a maximum mass $M_{\rm max}^{\rm dilute}$ and a
minimum mass $M_{\rm min}^{\rm dense}$. Eby {\it et
al.} \cite{ebycollapse,ebycollisions,ebychiral} studied the collapse of
dilute axion
stars to dense axion stars with the Gaussian ansatz\footnote{As
noted in Appendix B of \cite{phi6}, replacing a mathematical singularity (Dirac
peak) by a dense
axion star with a small radius does not change the estimate of the collapse time
obtained in
\cite{bectcoll}.} and argued that collapsing axion stars evaporate a large
fraction of their mass through the rapid emission of relativistic axions.

(ii) The second possibility is a {\it bosenova} phenomenon in which the collapse
of
the axion star may be accompanied by a burst of
outgoing relativistic axions
(radiation) produced by
inelastic reactions when the density reaches high values. In
that case, the collapse (implosion) is followed by an explosion. This
phenomenon
was shown 
experimentally by Donley {\it et al.} \cite{donley} for
nongravitational relativistic BECs with an attractive self-interaction and has
been
demonstrated
by Levkov {\it
et al.} \cite{tkachevprl} for
relativistic axion stars
 from direct
numerical
simulations of the KGE
equations in the Newtonian limit with the exact axionic potential taking
collisions into
account. These
equations
predict multiple cycles of collapses and explosions with a self-similar scaling
regime and a series of singularities at finite times. These multiple cycles can
lead either to a dilute axion star with a mass $M'<M_{\rm max}$ or {\it no
remnant} at all
because of complete disappearance of the axion star into scalar waves.

(iii) The third possibility, when the mass
of the axion star is sufficiently large or when the self-interaction is
sufficiently weak, is the formation of a  {\it black hole} \cite{helfer,moss}.
In
that case, general relativity must be taken into account. Helfer {\it et al.}
\cite{helfer} and Michel and Moss \cite{moss} produced a phase diagram
displaying a tricritical point joining phase boundaries between dilute axion
stars,
relativistic bosenova (no remnant), and black
holes.\footnote{Their phase diagram is consistent
with the maximum mass of nonrelativistic dilute axion stars with quartic
attractive self-interaction obtained in \cite{prd1,prd2} (see the solid line in
Fig. 3 of \cite{moss}).}

The importance of relativistic effects during the collapse of
axion stars has been stressed by Visinelli {\it et al.}
\cite{visinelli}. In particular, they argued that
special relativistic effects are crucial on the dense branch\footnote{Braaten
and Zhang \cite{braatenrevue} argue that their evidence is not completely
convincing except close to the minimum mass $M_{\rm min}^{\rm dense}$. The
accuracy of the
nonrelativistic approximation may improve as $M$ increases along the dense
branch.} while self-gravity can generally be neglected. As a result, dense axion
stars correspond to pseudo-breathers or oscillons which are described by the
sine-Gordon equation. These objects are known to be unstable
and to decay via emission of relativistic axions (more precisely, they
are dynamically stable but they decay
rapidly because of relativistic effects). They have a very short lifetime much
shorter than any cosmological timescale. Eby {\it et al.}
\cite{ebyexpansion,ebyclass,elssw} confirmed the claim
of Visinelli {\it et al.}
\cite{visinelli} that dense axion stars are relativistic and
short-lived.\footnote{By contrast, dilute axion stars are long-lived with
respect
to decay in photons with a lifetime
far longer than the age of the Universe
\cite{ebylifetime,ebydecay,ebybh,braatenR}. However, photons can be
emitted during collisions between dilute axion stars and neutron stars. In
particular,
it
has been proposed that fast radio bursts (FRBs), whose origin is one of the
major mysteries of high energy astrophysics,  could be caused by axion
stars that can engender bursts when undergoing conversion into photons during
their collision with the magnetosphere of neutron stars (magnetars),  during
their collision with
the
magnetized
accretion disk of a black hole, or
during their collapse above the maximum mass. We refer to
\cite{tkachev2015,iwazaki,raby,iwazakinew,bai} for the suggestion of this
scenario
and to
\cite{pshirkov} for an interesting critical discussion.} It is
important to stress that
these authors considered axions (like QCD axions) described by a {\it real}
scalar field for which the particle number is not conserved in the
relativistic regime. This is the reason for their fast decay. Alternatively,
if we consider ULAs described by a {\it complex} scalar field (like, e.g., in
Refs. \cite{abrilph,playa}) for which the particle number is conserved, the
dense axion stars should be long-lived. This is suggested by the recent work of
Guerra {\it et al.} \cite{guerra} on ``axion boson stars''.

Phase transitions between nonrelativistic  dilute and
dense axion stars have been studied in
\cite{phi6} using the Gaussian ansatz. This allowed us to recover analytically 
the mass-radius relation of axion stars obtained numerically in
\cite{braaten}.\footnote{In Ref. \cite{phi6} we have
argued that, at very large
masses where general relativistic effects are important, the mass-radius
relation of dense axion stars should form a spiral. This implies the existence
of another maximum mass $M_{\rm max,GR}^{\rm
dense}$, of general relativistic origin, 
above which the dense axion stars collapse towards a black hole. We have
estimated  this maximum mass qualitatively in \cite{phi6}. In this
manner, we could recover analytically \cite{phi6} 
the phase diagram and the tricritical point obtained numerically in Refs.
\cite{helfer,moss}.}
There exists a transition mass $M_t$ such
that
dilute axion stars are fully stable (global minima of energy) for $M<M_t$
and metastable (local  minima of energy) for $M_t<M<M_{\rm max}^{\rm dilute}$.
Inversely, dense axion stars are metastable for
$M_{\rm min}^{\rm dense}<M<M_t$ and fully stable for
$M>M_t$. If a dilute axion star gains mass, for instance by merger and
accretion, it can overcome the maximum mass $M_{\rm max}^{\rm dilute}$, collapse
and form a
dense axion star (it may also emit a relativistic radiation -- bosenova -- and
disappear into scalar waves as
discussed above). Inversely, if a dense axion star loses mass, decaying by
emitting axion
radiation because of relativistic effects, it can pass below
the minimum mass $M_{\rm min}^{\rm dense}$ and
disperse outwards
(explosion) due to the repulsive kinetic pressure (quantum
potential). This mechanism determines the
lifetime of dense axion stars in the nonrelativistic regime. As noted by Braaten
and Zhang \cite{braatenrevue} their lifetime may be too short to be
astrophysically relevant. However, dense axion stars may have an important
cosmological effect by transforming nonrelativistic axions into relativistic
axions. These phase
transitions, involving collapses and explosions, are similar to those studied in
\cite{ijmpb,calettre} for
self-gravitating fermions at finite temperature enclosed within a ``box''. They
also share similarities
with the phase transitions of compact objects (white dwarfs, neutron
stars and black holes) as discussed in Sec. XI.C of Ref. \cite{phi6}. This
analogy has been recently confirmed by Guerra {\it et
al.} \cite{guerra} who numerically
solved the KGE equations for a complex scalar field. Their mass-radius
relations display the Newtonian maximum mass of dilute axion stars $M_{\rm
max}^{\rm dilute}$ derived in \cite{prd1,prd2} and
the general relativistic maximum mass of dense axion stars $M_{\rm max,GR}^{\rm
dense}$ predicted
qualitatively in \cite{phi6} (see Appendix
\ref{sec_mgrdas} for a complementary discussion).

Close to the maximum mass $M_{\rm
max}^{\rm dilute}$, the  dilute axion stars are metastable (local but
not global minima of energy). They are rendered unstable  by the 
quantum mechanical process of barrier-penetration (tunnel effect). We can
determine the tunneling rate of axion stars, and their lifetime, by using the
theory of path integrals and instantons that was originally elaborated in the
context of quantum field theory \cite{coleman,cc}. The
instanton theory was
applied to the
Gross-Pitaevskii (GP) equation by Stoof \cite{stoof} in order to determine the
lifetime of a (nongravitational) metastable BEC with an attractive
self-interaction in a
confining harmonic potential.\footnote{Experimental evidence of
Bose-Einstein condensation was reported by several groups in 1995
\cite{aemwc,dmaddkk,bradley1}. Some laboratory BECs like 
$^7$Li are made of atoms that have a negative scattering length ($a_s<0$),
hence an attractive self-interaction \cite{bradley1}.  When
they are
confined by a harmonic potential, they
are stable
(actually metastable) only below a maximum particle number $N_{\rm max}$.
This maximum particle number was obtained by Ruprecht {\it et al.}
\cite{ruprecht} and Kagan {\it et al.}
\cite{kagan} by solving the GP equation numerically and by Baym and Pethick
\cite{bp} and Stoof \cite{stoof} by solving the GP
equation analytically using a Gaussian ansatz. The
approximate analytical approach
of Baym and Pethick \cite{bp} and Stoof \cite{stoof} --
called the method of collective coordinates or the Ritz optimization procedure
-- was further developed by \cite{perez1,perez2,sackett} and finds its
origin in the works of \cite{anderson,caglioti,desaix,aceves,rasmussen,michinel}
in the context of nonlinear optics. The existence of a maximum
particle number was confirmed
experimentally in Ref. \cite{bradley2}. Near the stability limit, quantum
tunneling or thermal fluctuations cause the condensate to collapse.
During the collapse, the density rises until collisions
cause atoms to be ejected from the condensate in an energetic explosion similar
to supernova \cite{hs}. After the explosion, the
condensate
regrows fed by collisions between thermal atoms in the gas. This leads to a
series of sawtooth-like cycles of growth (explosion) and collapse
\cite{ssh,kms,sgwh,gsph,sackett} until the gas reaches thermal equilibrium.}
Using
a Gaussian ansatz, he showed that this problem can
be reduced to the simpler problem of the quantum tunneling rate of a fictive
particle in a one dimensional
potential. This basically leads to the WKB formula
\cite{llquantique}. The approach of Stoof \cite{stoof}  was further developed
by
Ueda and Leggett \cite{leggett} and Huepe {\it et al.}
\cite{huepe} who studied the behavior of the tunneling rate close to the
critical point. The correctness of Stoof's analytical
approach was studied  by Freire and Arovas \cite{fa}  who used a more rigorous
instanton theory based on
field theory and
showed that the results of Stoof provide a relevant approximation of
the exact solution. We will assume that the collective coordinate
 approach (Gaussian ansatz) remains valid in the case of self-gravitating BECs
with an
attractive self-interaction and we will use this analytical
approach in line with our previous works on the subject
\cite{prd1,prd2,bectcoll,phi6,epjpbh}. A similar investigation was recently made
by Eby
{\it et al.} \cite{ebybh}.
Here,
we explicitly derive the analytical
expression
of the quantum tunneling rate close to the maximum mass emphasizing the scaling
$(1-M/M_{\rm max})^{5/4}$ of the reduction factor. Despite
this reduction
factor, we show
that
the lifetime of
metastable axion
stars is
considerable, scaling as $e^N t_D$ (where $t_D$
is the dynamical time)
with $N\sim 10^{57}$ and $t_D\sim 10\, {\rm hrs}$ for QCD
axions
and
$N\sim 10^{96}$ and $t_D\sim 100\, {\rm
Myrs}$ for ULAs.\footnote{The scaling $e^N t_D$  was anticipated in
Ref. \cite{phi6} by analogy with similar results obtained for other systems with
long-range interactions, such as globular clusters \cite{lifetime}, where the
destabilization of the metastable state is due to thermal (or
energetical) fluctuations instead of quantum fluctuations.} Therefore, in
practice, metastable
states can be
considered as
stable equilibrium states, except for masses extraordinarily close to the
maximum mass $M_{\rm max}$. We develop a finite size scaling
theory close to the maximum mass and
predict that the collapse time at criticality scales as $t_{\rm
coll}\sim N^{1/5}t_D$ instead of
being infinite as in Ref. \cite{bectcoll} where fluctuations
are neglected. The collapse time
is smaller than the age of the universe for QCD axion stars and larger for
ULAs. On the other hand, our detailed calculation of the quantum tunneling
rate may be
useful
if one is able in the future to perform direct $N$-body simulations or
laboratory experiments of self-gravitating BECs with an attractive
self-interaction mimicking dilute axion stars. In that case, the number of
bosons $N$ will not be
very large and metastability effects should be observed, especially close to
the maximum mass.

This paper is organized as follows. In Sec. \ref{sec_gbm} we recall the basic
equations describing dilute axion stars. In Sec. \ref{sec_ga} we use a Gaussian
ansatz to transform these equations into the simpler mechanical problem of a
fictive particle in a one dimensional potential. In Sec. \ref{sec_qtr} we 
determine the quantum tunneling rate of the BEC from the theory of instantons.
We give its general expression and its approximate expression close to the
maximum mass. In Sec. \ref{sec_ttr} we briefly consider the thermal tunneling
(or thermal activation) rate of the BEC by using the analogy with Brownian
motion. In Sec. \ref{sec_k}
we consider corrections to the maximum mass due to
quantum and thermal
fluctuations and show that they are generally negligible. We emphasize the
very long lifetime of dilute axion stars.  Finally, in Sec.
\ref{sec_size} we determine the correction to the collapse time at
criticality due to quantum and thermal fluctuations. We finally conclude by
discussing analogies and differences with other systems of physical interest.

\section{Dilute axion stars}
\label{sec_gbm}

In this section, we recall the basic equations describing dilute axion stars in
the nonrelativistic limit.

\subsection{GPP equations}
\label{sec_gpp}

Dilute axion stars can be interpreted as Newtonian self-gravitating BECs with an
attractive self-interaction. They are described by the GPP
equations\footnote{See Appendix \ref{sec_kge} for the derivation of the GPP
equations from the more general KGE equations describing relativistic
axion
stars.} 
\begin{eqnarray}
\label{gpp1}
i\hbar \frac{\partial\psi}{\partial
t}=-\frac{\hbar^2}{2m}\Delta\psi+m\Phi\psi+\frac{4\pi
a_s\hbar^2}{m^2}|\psi|^{2}\psi,
\end{eqnarray}
\begin{equation}
\label{gpp2}
\Delta\Phi=4\pi G |\psi|^2,
\end{equation}
where $\psi({\bf r},t)$ is the wave function of the condensate, $\Phi({\bf
r},t)$ is the gravitational potential, and $a_s$ is the scattering length of the
bosons ($a_s<0$ for axions with an attractive self-interaction). The GP
equation (\ref{gpp1}) involves a cubic nonlinearity associated with a quartic
effective potential (see Eq. (\ref{kge27}) of Appendix  \ref{sec_kge}).

\subsection{Hydrodynamic equations}
\label{sec_he}

Making the  Madelung \cite{madelung} transformation
\begin{equation}
\label{he1}
\psi({\bf r},t)=\sqrt{{\rho({\bf r},t)}} e^{iS({\bf r},t)/\hbar},\quad
\rho=|\psi|^2,\quad {\bf u}=\frac{\nabla S}{m},
\end{equation}
where $\rho({\bf r},t)$ is the mass density, $S({\bf r},t)$ is the action
and
${\bf u}({\bf r},t)$ is the velocity field, it can be shown
(see, e.g., \cite{chavtotal})
that
the GPP equations (\ref{gpp1}) and (\ref{gpp2})  are equivalent to
hydrodynamic
equations of the
form
\begin{equation}
\label{he2}
\frac{\partial\rho}{\partial t}+\nabla\cdot (\rho {\bf u})=0,
\end{equation}
\begin{equation}
\label{he3}
\frac{\partial {\bf u}}{\partial t}+({\bf u}\cdot \nabla){\bf
u}=-\frac{1}{\rho}\nabla P-\nabla\Phi-\frac{1}{m}\nabla
Q,
\end{equation}
\begin{equation}
\label{he4}
\Delta\Phi=4\pi G\rho,
\end{equation}
where
\begin{equation}
\label{he5}
Q=-\frac{\hbar^2}{2m}\frac{\Delta
\sqrt{\rho}}{\sqrt{\rho}}=-\frac{\hbar^2}{4m}\left\lbrack
\frac{\Delta\rho}{\rho}-\frac{1}{2}\frac{(\nabla\rho)^2}{\rho^2}\right\rbrack
\end{equation}
is the quantum potential taking into account the Heisenberg uncertainty
principle and $P(\rho)$ is the pressure arising from the self-interaction
of the bosons. For a cubic nonlinearity (i.e. a $|\psi|^4$ effective potential),
the
equation of state is quadratic
\begin{equation}
\label{he6}
P=\frac{2\pi a_s\hbar^2}{m^3}\rho^{2}.
\end{equation}
This is a  polytropic equation of state of index
$n=1$. For an attractive self-interaction between the bosons ($a_s<0$), the
pressure is negative. Equations
(\ref{he2})-(\ref{he4}) are called the quantum Euler-Poisson equations. They are
equivalent to the GPP equations (\ref{gpp1}) and  (\ref{gpp2}).
In the following, we will exclusively use the
hydrodynamic formalism. In that case, the normalization condition of the
wave function is equivalent to
the conservation of mass $M=\int \rho\, d{\bf r}$.
We refer to \cite{chavtotal}
for the expression of the following results in terms of the wave function.

\subsection{Equilibrium state}
\label{sec_eq}

In the hydrodynamic representation, an equilibrium state of the quantum
Euler-Poisson equations (\ref{he2})-(\ref{he4}), obtained
by taking $\partial_t=0$
and ${\bf u}={\bf 0}$, satisfies the equation
\begin{equation}
\label{eq1}
\nabla P+\rho\nabla\Phi+\frac{\rho}{m}\nabla Q={\bf
0}.
\end{equation}
This equation can be interpreted as a condition of quantum hydrostatic
equilibrium. It describes the balance between the pressure
due to the self-interaction of the bosons, the gravitational force, and the
quantum force arising from the Heisenberg
uncertainty principle.  Combining Eq. (\ref{eq1}) with the Poisson
equation (\ref{he4}), we
obtain the fundamental differential equation of quantum hydrostatic equilibrium
\cite{chavtotal}
\begin{equation}
\label{eq2}
-\nabla\cdot \left (\frac{\nabla P}{\rho}\right )+\frac{\hbar^2}{2m^2}\Delta
\left (\frac{\Delta\sqrt{\rho}}{\sqrt{\rho}}\right )=4\pi G\rho.
\end{equation}
For the quadratic equation of state (\ref{he6}), this differential equation has
been solved numerically in Ref. \cite{prd2} in the general case of attractive or
repulsive self-interaction.

\subsection{Total energy}
\label{sec_te}

The total energy associated with the quantum
Euler-Poisson equations is given by
\begin{eqnarray}
\label{te1}
E_{\rm tot}=\Theta_c+\Theta_Q+U+W,
\end{eqnarray}
where $\Theta_c$ is the classical kinetic energy, $\Theta_Q$ is the quantum
kinetic energy, $U$ is the internal energy, and $W$ is the gravitational energy.
It can be explicitly written as
\begin{eqnarray}
\label{te2}
E_{\rm tot}&=&\int\rho \frac{{\bf u}^2}{2}\, d{\bf r}+\frac{1}{m}\int \rho Q\,
d{\bf r}\nonumber\\
&+&\frac{2\pi a_s\hbar^2}{m^3}\int \rho^2\, d{\bf r}
+\frac{1}{2}\int\rho\Phi\, d{\bf r}.
\end{eqnarray}
We can easily show \cite{chavtotal} that the quantum Euler-Poisson  equations
(\ref{he2})-(\ref{he4}) conserve the total energy ($\dot E_{\rm tot}=0$).

\subsection{Variational principle}
\label{sec_neq}

It can be shown
that the minimization problem
\begin{eqnarray}
\label{neq1}
\min_{\rho,{\bf u}} \left\lbrace E_{\rm tot}[\rho,{\bf u}]\quad |\quad
M\quad {\rm
fixed}\right\rbrace.
\end{eqnarray}
determines an equilibrium state of the quantum Euler-Poisson equations that is
dynamically stable. This
is a criterion of nonlinear
dynamical
stability resulting from the fact that $E_{\rm tot}$ and $M$ are conserved by 
the quantum Euler-Poisson equations. It provides a necessary and
sufficient condition of dynamical stability since it takes into
account all the invariants of the quantum Euler-Poisson equations.

The
variational principle for the first variations
(extremization) can be written as 
\begin{eqnarray}
\label{neq2}
\delta E_{\rm tot}-\frac{\mu}{m}\delta M=0,
\end{eqnarray}
where $\mu$ is a Lagrange multiplier (chemical potential) taking into account
the mass constraint. This variational problem gives
${\bf u}={\bf 0}$ (the equilibrium state is static) and the Gibbs condition 
\begin{eqnarray}
\label{neq3}
m\Phi+\frac{4\pi a_s\hbar^2}{m^2}\rho+Q=\mu.
\end{eqnarray}
Taking the gradient of Eq.
(\ref{neq3}) and using Eq. (\ref{he6}), we recover the condition of  quantum
hydrostatic equilibrium
(\ref{eq1}). Therefore, an extremum of total energy at fixed mass is a steady
state of the
quantum Euler-Poisson equations. Furthermore, it can be shown
that the star
is linearly stable with respect to the quantum Euler-Poisson equations if, and
only if,
it is a local minimum of energy  at fixed mass (a maximum or a saddle
point is linearly unstable).

Using
the Poincar\'e criterion \cite{poincare} or the catastrophe (or bifurcation)
theory \cite{catastrophe},\footnote{The  Poincar\'e turning point criterion
\cite{poincare} states
that a mode of stability is lost at an extremum of mass if the curve $\mu(M)$
rotates anticlockwise and gained if it rotates clockwise. It
is equivalent to the mass-radius theorem of Wheeler \cite{htww} introduced in
the physics of compact objects like white dwarfs and neutron stars. It states
that a mode of stability is lost at an extremum of mass if the curve $M(R)$
rotates anticlockwise and gained if it rotates clockwise. To be
complete, we also
quote the  necessary Vakhitov-Kolokolov condition of stability $dM/d\rho>0$
\cite{zakharov,vk}.}  we can generically conclude that the
series of equilibria is dynamically stable before
the turning points of mass $M$ or energy $E_{\rm tot}$ (they coincide) and that
it
becomes
dynamically unstable
afterwards.\footnote{Note that
in certain situations a mode of stability can
be regained after a turning point of mass. We refer to
\cite{katzpoincare,htww,ijmpb} for a detailed account of the Poincar\'e
criterion and of  Wheeler's $M(R)$ theorem when there are multiple turning
points.} Furthermore, the
curve $E_{\rm tot}(M)$ displays cusps at its extremal points (since
$\delta E_{\rm tot}=0\Leftrightarrow \delta M=0$).

\subsection{Quantum virial theorem}
\label{sec_qvt}

The time-dependent scalar  virial theorem associated with the quantum
Euler-Poisson equations can be written as (see Appendix G of \cite{chavtotal})
\begin{equation}
\label{qvt1}
\frac{1}{2}\ddot I=2(\Theta_c+\Theta_Q)+3\int
P\, d{\bf r}+W,
\end{equation}
where $I=\int\rho r^2\, d{\bf r}$ is the moment of inertia.
At equilibrium, we obtain the quantum virial theorem
\begin{equation}
\label{qvt2}
2\Theta_Q+3\int P\, d{\bf r}+W=0.
\end{equation}

\section{Gaussian ansatz}
\label{sec_ga}

\subsection{Total energy}
\label{sec_gte}

We can obtain an approximate analytical solution
of the GPP equations (\ref{gpp1}) and (\ref{gpp2}) by
developing a mechanical analogy. Making a Gaussian ansatz for the wavefunction
(see, e.g.,  Sec. 8.2 of Ref. \cite{chavtotal} for details):
\begin{eqnarray}
\label{gte1a}
\psi({\bf r},t)=\left \lbrack \frac{M}{\pi^{3/2}
R(t)^3}\right \rbrack^{1/2}e^{-r^2/2R(t)^2}e^{imH(t)r^2/2\hbar},
\end{eqnarray}
where $R(t)$ is the typical
radius of the BEC and $H=\dot R/R$, we find that the
energy
functional (\ref{te1}) 
can be written
as a function of $R$ and $\dot R$  (for a fixed mass $M$) as\footnote{For a
Gaussian density profile, the relation
between the radius $R$ and the
radius $R_{99}$ containing $99\%$
of the mass is $R_{99}=2.38167 R$  \cite{prd1}.}
\begin{eqnarray}
\label{gte1}
E_{\rm tot}=\frac{1}{2}\alpha M\left (\frac{dR}{dt}\right )^2+V(R)
\end{eqnarray}
with the effective potential
\begin{eqnarray}
\label{gte2}
V(R)=\sigma \frac{\hbar^2M}{m^2R^2}-\zeta\frac{2\pi
|a_s|\hbar^2M^2}{m^3R^{3}}-\nu
\frac{GM^2}{R}.
\end{eqnarray}
The coefficients are
\begin{eqnarray}
\label{gte2b}
\alpha=\frac{3}{2},\quad \sigma=\frac{3}{4},  \quad
\zeta=\frac{1}{(2\pi)^{3/2}}, \quad \nu=\frac{1}{\sqrt{2\pi}}.
\end{eqnarray}
The first term in Eq. (\ref{gte1}) is the classical kinetic energy while the  
effective potential (\ref{gte2}) comprises the quantum kinetic energy,
the internal energy and the gravitational energy. Using the conservation of
total energy, $\dot E_{\rm tot}=0$, we get
\begin{eqnarray}
\label{gte3}
\alpha M\frac{d^2R}{dt^2}=-\frac{d{V}}{dR}.
\end{eqnarray}
This equation is similar to the equation of motion of a particle of
mass $\alpha M$ and position $R$ moving in a one-dimensional potential $V(R)$.
This equation can also be obtained from the quantum virial theorem
(\ref{qvt1}) (see Sec. 8.4 of Ref.
\cite{chavtotal} for details). Instead of starting from the total
energy, the same results can be obtained from the Lagrangian of the GPP
equations (see Appendix B of Ref. \cite{bectcoll} for details). Finally, we can
draw some analogies between the equation of motion (\ref{gte3}) for the radius
of a BEC and the Friedmann equations in cosmology governing the evolution of the
scale factor of the Universe where $H=\dot R/R$ plays the role of the Hubble
constant (see Sec. 8.8 of Ref. \cite{chavtotal} for
details).

\subsection{Mass-radius relation}
\label{sec_mr}

\begin{figure}[!h]
\begin{center}
\includegraphics[clip,scale=0.3]{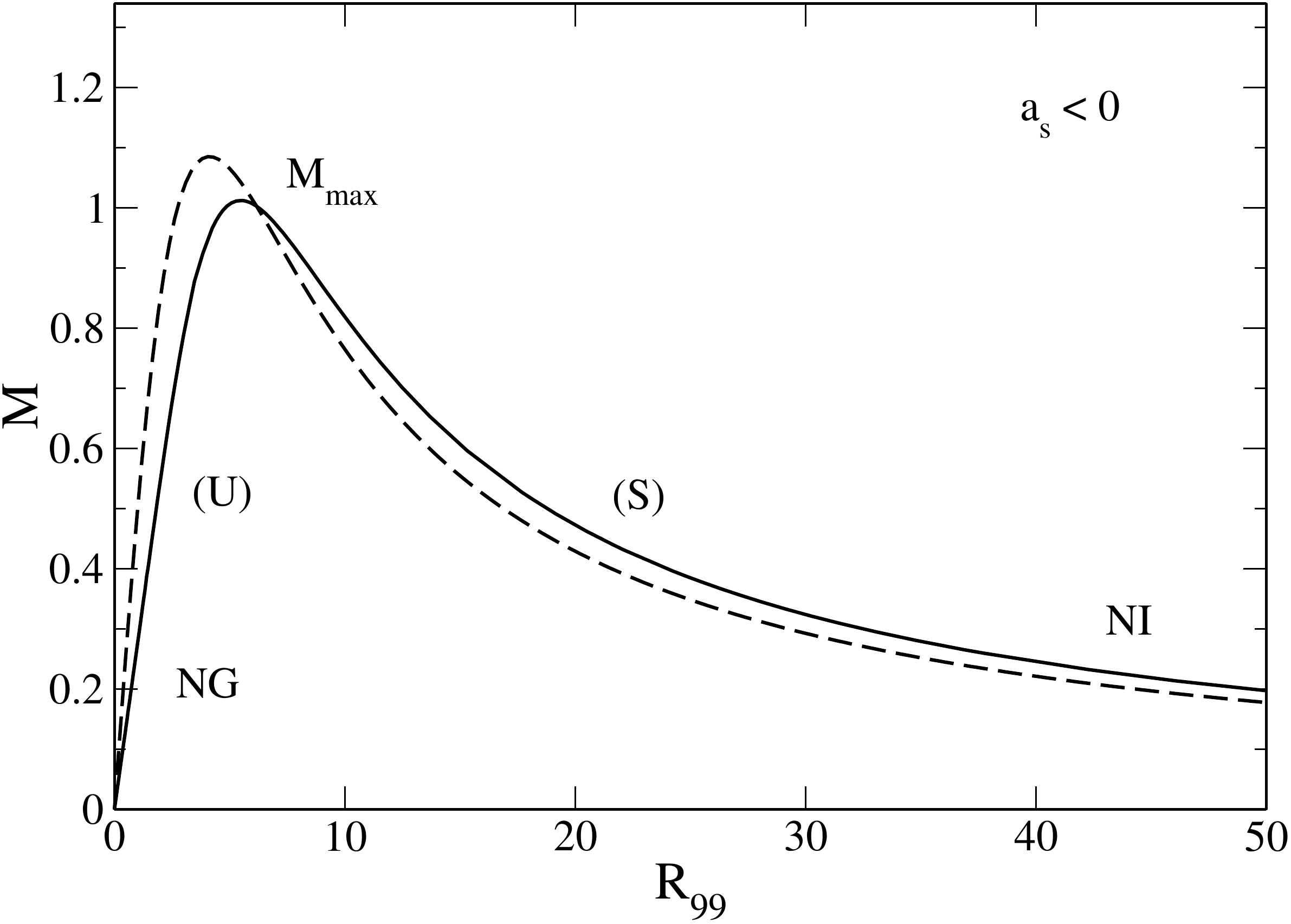}
\caption{Mass-radius relation of dilute
axion stars interpreted as a self-gravitating BEC with an attractive
self-interaction ($a_s<0$). We have chosen a normalization such that
$\hbar=G=m=|a_s|=1$. The solid line is the exact mass-radius relation obtained
by solving the GPP equations numerically \cite{prd2}. The dotted line
corresponds to the approximate analytical mass-radius relation 
(\ref{mr1}) obtained from the Gaussian ansatz \cite{prd1}.}
\label{MRmodif}
\end{center}
\end{figure}

We have seen that an extremum of total energy $E_{\rm tot}$ given by Eq.
(\ref{te1}) at fixed mass $M$ is an equilibrium state of the GPP equations
(\ref{gpp1}) and (\ref{gpp2}). On the other hand,  a (local)
minimum of total energy is (meta)stable while a maximum or a saddle point is
unstable. Within the Gaussian ansatz, we have to minimize the 
total energy $E_{\rm tot}$ given by Eq. (\ref{gte1}) at fixed mass $M$. An
extremum
corresponds to $dR/dt=0$ and $V'(R)=0$. The second condition leads
to
the mass-radius relation \cite{prd1}
\begin{eqnarray}
\label{mr1}
M=\frac{2\sigma\frac{\hbar^2}{m^2R^3}}{\frac{\nu
G}{R^2}+6\pi\zeta\frac{|a_s|\hbar^2}{m^3R^4}}.
\end{eqnarray}
This relation  is plotted in Fig. \ref{MRmodif} in the case of an
attractive self-interaction ($a_s<0$). It displays a 
maximum mass
\cite{prd1}  
\begin{eqnarray}
\label{mr2a}
M_{\rm max}=\left (\frac{\sigma^2}{6\pi\zeta\nu}\right
)^{1/2}\frac{\hbar}{\sqrt{Gm|a_s|}}
\end{eqnarray}
at
\begin{eqnarray}
\label{mr2b}
R_{*}=\left
(\frac{6\pi\zeta}{\nu}\right )^{1/2}\left
(\frac{|a_s|\hbar^2}{Gm^3}\right )^{1/2}.
\end{eqnarray}
The prefactors are $1.085$ and $1.73$. We have the identity
\begin{eqnarray}
M_{\rm max}=\frac{\sigma}{\nu}\frac{\hbar^2}{Gm^2 R_*}.
\label{mr3}
\end{eqnarray}
There is no equilibrium state with $M>M_{\rm max}$. When $M<M_{\rm max}$, two
equilibrium states exist with the same mass. By
computing the second
derivative of $V(R)$ or by using  the identity
\begin{eqnarray}
\frac{dM}{dR}=-\alpha M\frac{m^2R^3}{2\sigma \hbar^2}\omega^2,
\label{mr3b}
\end{eqnarray}
where $\omega^2=V''(R)/\alpha M$ is the square radial pulsation of the BEC (see
Sec. 8 of Ref. \cite{chavtotal} for details and generalizations)
one can analytically show \cite{prd1} that the equilibrium
states with $R>R_*$ are stable while the equilibrium
states with $R<R_*$ are unstable. Therefore, $R_*$ is the minimum radius of
stable equilibrium states. This result can also be obtained from the
mass-radius relation $M(R)$ by using the Poincar\'e turning point criterion
\cite{poincare} or the Wheeler theorem (see footnote 13) stating that
the change of stability occurs at the turning point
of mass (this result is valid beyond the Gaussian
ansatz as discussed in Sec. \ref{sec_neq}).

In the nongravitational limit, corresponding to $R\ll R_*$, the mass-radius
relation reduces to
\begin{eqnarray}
\label{mr4}
M\sim \frac{\sigma}{3\pi\zeta}\frac{mR}{|a_s|}.
\end{eqnarray}
However, all these equilibrium states are unstable. In the noninteracting limit,
corresponding to $R\gg R_*$, we get
\begin{eqnarray}
\label{mr5}
M\sim \frac{2\sigma}{\nu}\frac{\hbar^2}{Gm^2R}.
\end{eqnarray}
These equilibrium states are stable.

{\it Remark:} The above results apply to dilute axion stars. Stable  dilute
axion stars exist
only below a maximum mass $M_{\rm max}$ and above a minimum radius
$R_{*}$ given by Eqs. (\ref{mr2a}) and (\ref{mr2b}) within the Gaussian ansatz
\cite{prd1}. The
exact values of the maximum mass
$M_{\rm max}^{\rm
exact}$ and of the
corresponding radius $(R_{99}^*)^{\rm exact}$ [see Eqs.
(\ref{intro1a}) and (\ref{intro1b})] have been obtained in Ref. \cite{prd2} by
computing the steady states of the GPP equations (\ref{gpp1})
and (\ref{gpp2})
numerically. If we take into account a $\varphi^6$ repulsion in the potential of
self-interaction (or consider the exact potential of axions), an additional
stable branch appears in the mass-radius relation at small radii corresponding
to dense axion stars
\cite{braaten,phi6}.

\subsection{Collapse, gravitational cooling, or explosion}
\label{sec_ce}

When $M<M_{\rm max}$ there are two possible equilibrium states for the same
mass with radius $R_S>R_*$ and $R_U<R_*$. The equilibrium state $R_S$ is stable
(S) and the equilibrium state $R_U$ is unstable (U). The evolution of the
unstable state depends on the sign of its energy $E_{\rm tot}^{(U)}$. In
\cite{bectcoll} we have identified another critical mass 
\begin{eqnarray}
\label{ce1}
M_c=\frac{\sqrt{3}}{2}M_{\rm max}
\end{eqnarray}
at which $E_{\rm tot}^{(U)}=0$.

\begin{figure}
\begin{center}
\includegraphics[clip,scale=0.3]{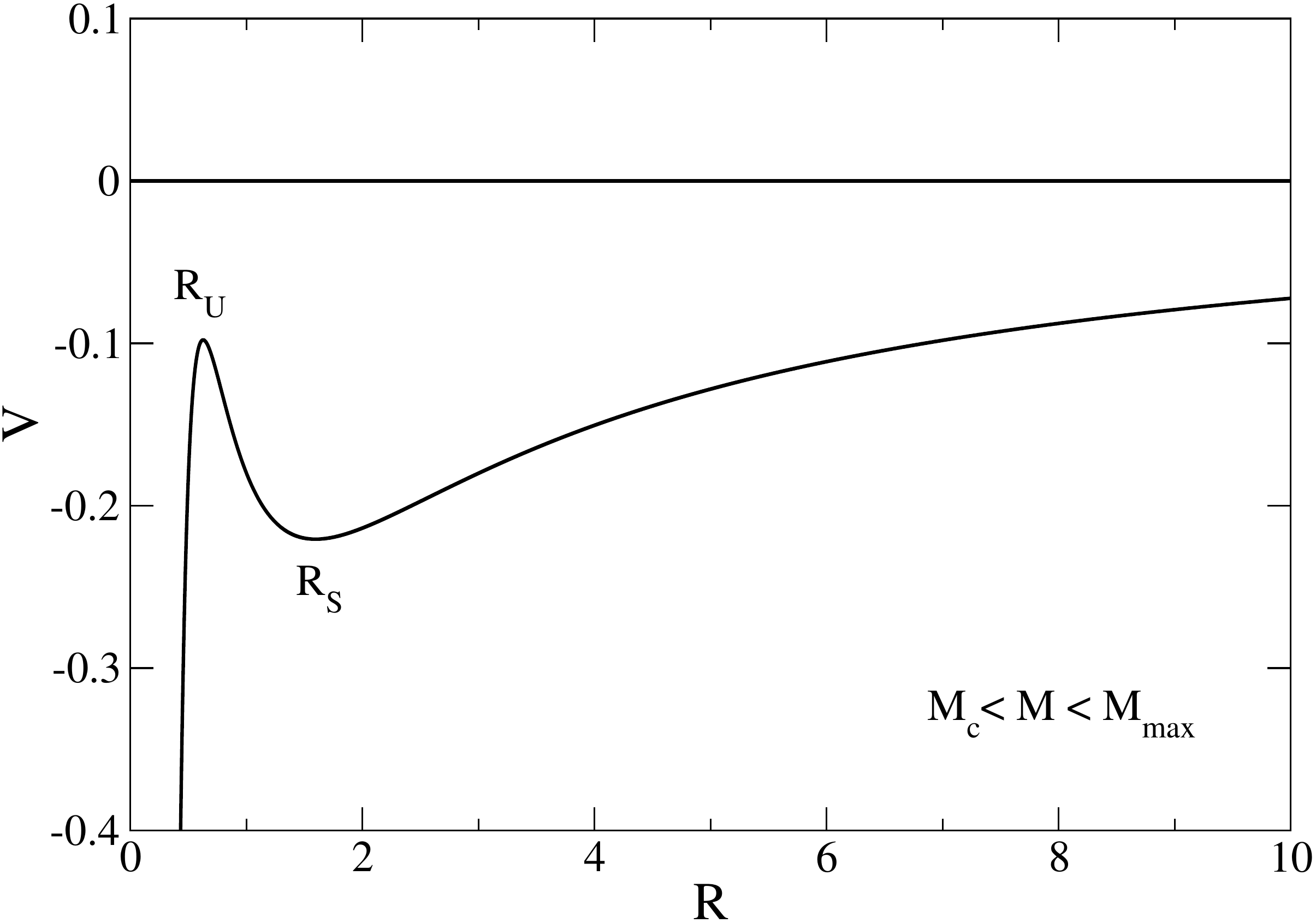}
\caption{Effective potential $V(R)$ as a function of the radius $R$ for
$M_c< M<M_{\rm max}$. In that case $V_{\rm max}<0$.}
\label{rvM0p9}
\end{center}
\end{figure}

When $M_c<M<M_{\rm max}$, the energy $E_{\rm tot}^{(U)}$ of the unstable state
is negative (see Fig. \ref{rvM0p9}). If slightly perturbed, the unstable star
can either collapse towards a Dirac peak  ($R\rightarrow 0$)
or migrate towards a stable dilute axion star  by
gravitational cooling ($R\rightarrow R_S$) \cite{seidel94,gul0,gu}. This is a
dissipative process similar to violent relaxation \cite{lb} during which the
star undergoes damped oscillations and emits a scalar field radiation. Through
this process, it loses energy (and mass) and settles on a stable equilibrium
state (S) with a larger radius and a lower energy than the initial configuration
(U).

\begin{figure}
\begin{center}
\includegraphics[clip,scale=0.3]{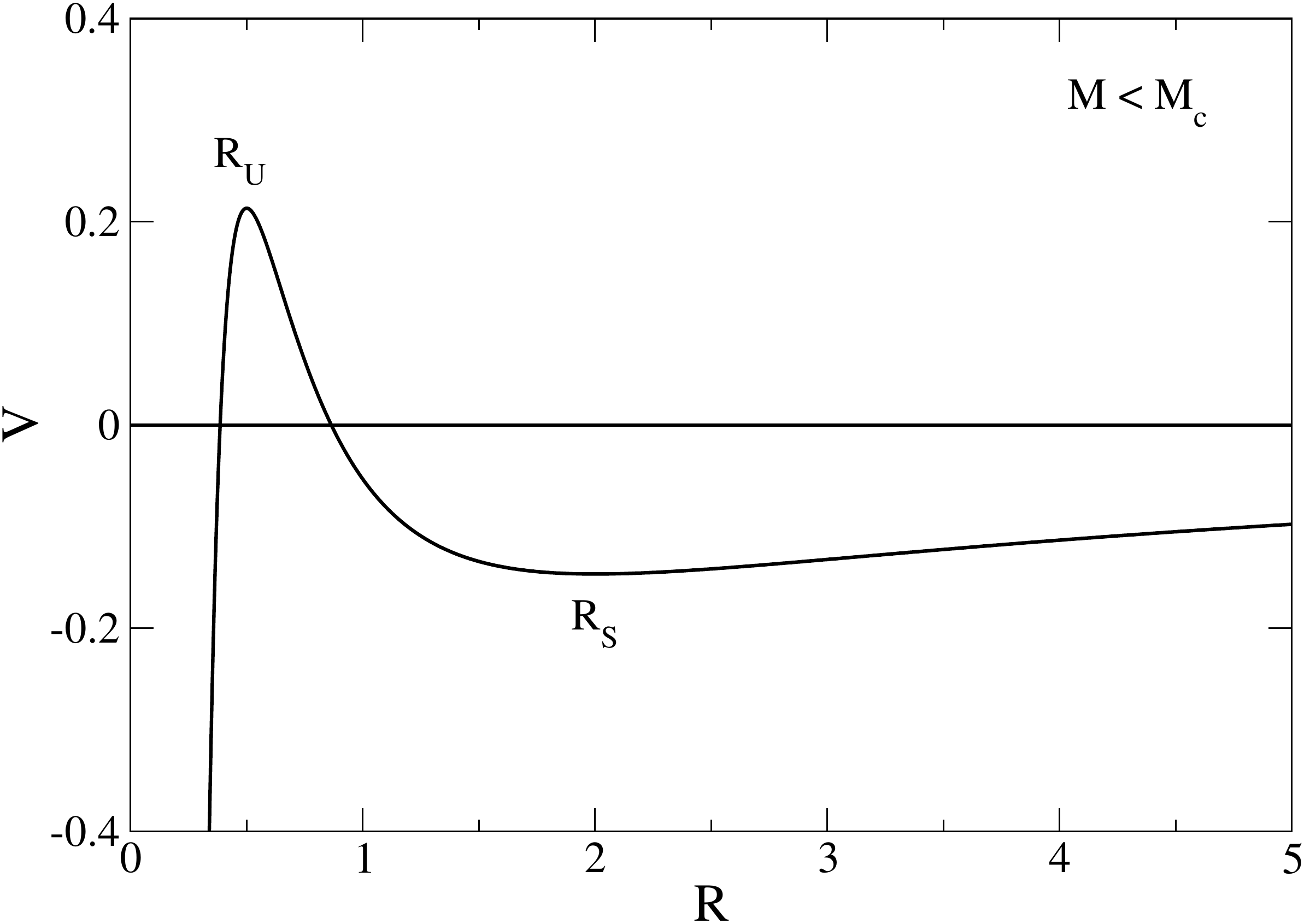}
\caption{Effective potential $V(R)$ as a function of the radius $R$ for
$M<M_{c}$. In that case $V_{\rm max}>0$.}
\label{rvM0p8}
\end{center}
\end{figure}

When $M<M_{c}$, the energy $E_{\rm tot}^{(U)}$ of the unstable state
is positive (see Fig. \ref{rvM0p8}). If slightly perturbed, the unstable star
can either collapse towards
a Dirac peak ($R\rightarrow 0$), migrate towards a stable dilute
axion star by gravitational cooling ($R\rightarrow R_S$), or
explode and disperse away ($R\rightarrow +\infty$). 

{\it Remark:} In the following, we shall assume that the mass of
the dilute
axion star (S) is
relatively close to $M_{\rm max}$. As a result, if it can reach the unstable
state (U) by quantum or thermal tunneling, thereby reducing its radius, it then
generically collapses towards the
Dirac peak.

\subsection{Normal form of the potential close to the maximum mass}
\label{sec_nfcmm}

\begin{figure}
\begin{center}
\includegraphics[clip,scale=0.3]{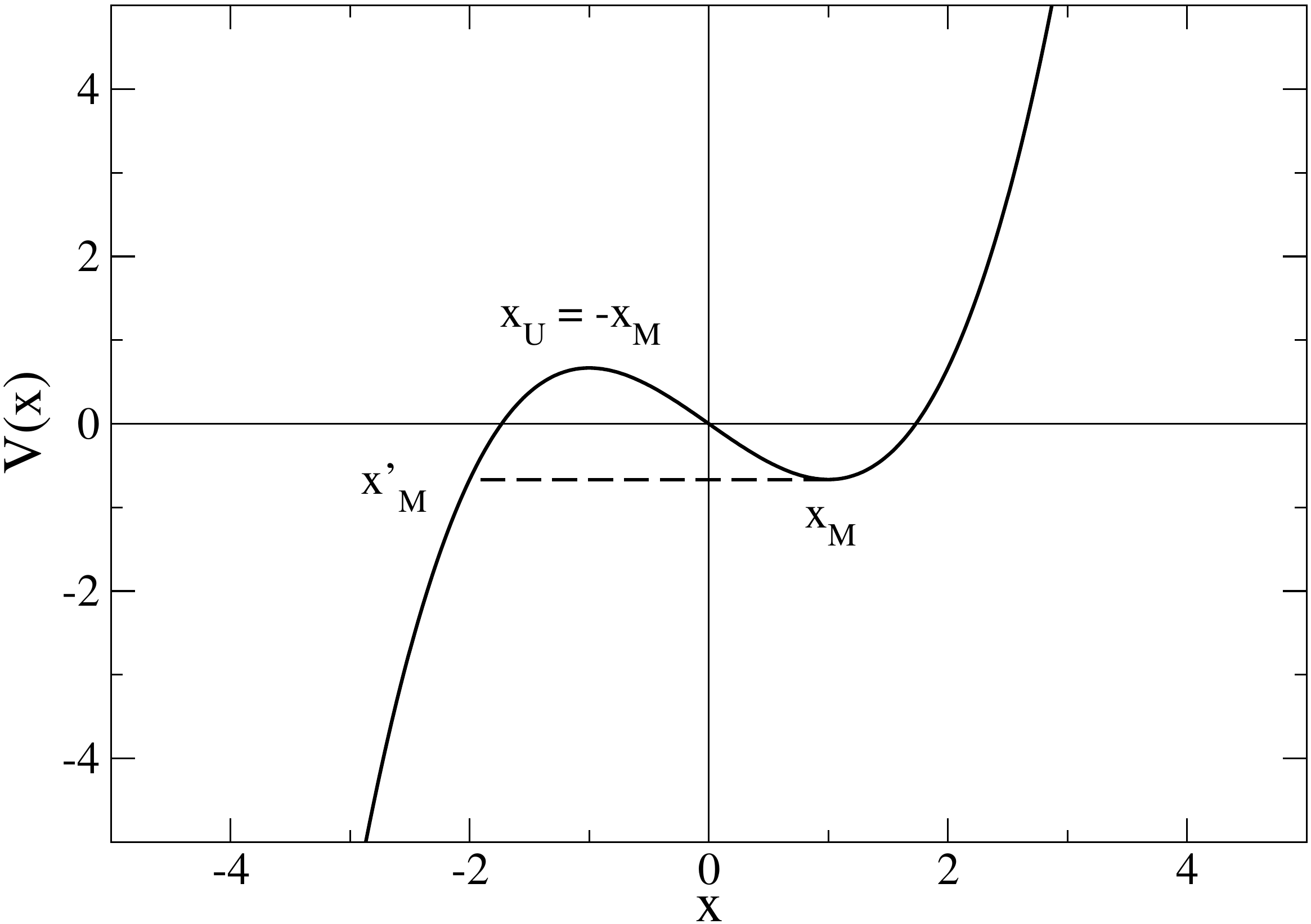}
\caption{Normal form of the potential close to a saddle-center
bifurcation.}
\label{normalform}
\end{center}
\end{figure}

Expanding the effective potential from Eq. (\ref{gte2}) to third order
close to the maximum mass $M_{\rm max}$, we obtain \cite{bectcoll}
\begin{eqnarray}
\label{nfcmm1}
\frac{V(R)}{V_0}&=&\frac{1}{3R_*^3}(R-R_*)^3-\frac{2}{R_*}\left
(1-\frac{M}{M_{\rm max}}\right )(R-R_*)\nonumber\\
&-&\frac{1}{3}+\frac{5}{3}
\left (1-\frac{M}{M_{\rm max}}\right ),
\end{eqnarray}
where
\begin{eqnarray}
V_0=\nu \frac{GM_{\rm
max}^2}{R_*}=\frac{\sigma^2\nu^{1/2}}{(6\pi\zeta)^{3/2}}\frac{\hbar
m^{1/2}G^{1/2}}{|a_s|^{3/2}}.
\label{nfcmm2}
\end{eqnarray}
Equation (\ref{nfcmm1}) is the normal form of a potential $V(R)$
close to a saddle-center bifurcation (see Fig. \ref{normalform}). With this
approximation, the equation of motion (\ref{gte3}) of the fictive particle
becomes
\begin{eqnarray}
\alpha M
\frac{d^2R}{dt^2}=-\frac{V_0}{R_*^3}(R-R_*)^2+\frac{2V_0}{R_*}\left
(1-\frac{M}{M_{\rm max}}\right ).
\end{eqnarray}

The mass-radius
relation close to  $M_{\rm max}$, corresponding to $V'(R)=0$, is
given by
\begin{eqnarray}
R-R_*=\pm \sqrt{2}R_*\left (1-\frac{M}{M_{\rm max}}\right )^{1/2}.
\label{nfcmm3}
\end{eqnarray}
The upper sign corresponds to the branch $R>R_*$ and the lower
sign corresponds to the branch $R<R_*$.  
On the other hand, the square radial pulsation 
$\omega^2=V''(R)/\alpha M$ of the BEC
is given by
\begin{eqnarray}
\omega^2=\pm \frac{2\sqrt{2}}{t_D^2} \left (1-\frac{M}{M_{\rm max}}\right
)^{1/2},
\label{nfcmm4}
\end{eqnarray}
where we have introduced the dynamical time
\begin{eqnarray}
t_D=\left (\frac{\alpha}{\nu}\right
)^{1/2}\frac{1}{\sqrt{G\rho_0}}=\frac{6\pi\zeta}{\nu}\left
(\frac{\alpha}{\sigma}\right
)^{1/2}\frac{|a_s|\hbar}{Gm^2}
\label{nfcmm5}
\end{eqnarray}
constructed with the density
\begin{eqnarray}
\label{nfcmm6}
\rho_{0}=\frac{M_{\rm max}}{R_{*}^3}=\frac{\sigma
\nu}{(6\pi\zeta)^2}\frac{Gm^4}{a_s^2\hbar^2}.
\end{eqnarray}
The expression from Eq. (\ref{nfcmm4}) confirms that the branch $R>R_*$ is
stable
($\omega^2>0$) while the branch $R<R_*$ is unstable ($\omega^2<0$). From Eqs.
(\ref{nfcmm3}) and (\ref{nfcmm4}) we obtain 
the relation
\begin{eqnarray}
\label{nfcmm7}
\frac{dM}{dR}=-\frac{\omega^2}{2}\frac{M_{\rm max}t_D^2}{R_*}
\end{eqnarray}
that links the stability of the system (through the sign of the square
pulsation $\omega^2$) to the slope of the mass-radius relation. This is a
particular case of the Poincar\'e turning point criterion close to the maximum
mass (see Sec. 8.7 of Ref. \cite{chavtotal} for
details). Another manner to investigate the
stability of an equilibrium state is to compute its energy. The energy of the
equilibrium states close to the maximum mass are
\begin{eqnarray}
\label{nfcmm8}
\frac{E_{\rm tot}}{V_0}=\frac{V}{V_0}=-\frac{1}{3}+\frac{5}{3}\left
(1-\frac{M}{M_{\rm max}}\right
)\nonumber\\
\mp\frac{4}{3}\sqrt{2}\left (1-\frac{M}{M_{\rm max}}\right
)^{3/2}.
\end{eqnarray}
As expected, the energy of the stable state ($R>R_*$, upper sign) is lower
than the energy of the unstable state ($R<R_*$, lower sign) for the same mass
$M$.

When $M>M_{\rm max}$, there is no equilibrium state. In that case, the
dilute axion star is expected to collapse. 
Within our approximations (nonrelativistic treatment $+$ purely attractive
self-interaction $+$ spherical collapse), it should form a classical singularity
(Dirac peak). The collapse time has been investigated in  \cite{bectcoll} using
a Gaussian ansatz. It is found that, close to the maximum mass, the
collapse time is given by
\begin{equation}
\label{nfcmm9}
\frac{t_{\rm coll}}{t_D}\sim 2.90178... \left (\frac{M}{M_{\rm max}}-1\right
)^{-1/4} \quad (M\rightarrow M_{\rm max}^+).
\end{equation}
If we consider that the dilute axion star  collapses towards a
dense axion star of finite radius $R_{\rm dense}>0$
\cite{ebycollapse,phi6} instead of forming a singularity at $R=0$ (Dirac
peak), the results
of \cite{bectcoll} remain valid because $R_{\rm dense}$ is generically very
small (this point is specifically addressed in Appendix B of \cite{phi6}).

\section{Quantum tunneling rate of the BEC}
\label{sec_qtr}

When $M<M_{\rm max}$, the potential $V(R)$ has two equilibrium states (see Fig.
\ref{normalform}): a
stable equilibrium state at $R_M>R_*$ (local minimum) and an unstable
equilibrium state at $R_U<R_*$ (local maximum). Since the potential 
 $V(R)$ has no global minimum (it tends to $-\infty$ when $R\rightarrow 0$),
the stable equilibrium state at $R_M$ is actually metastable. This
metaequilibrium state represents a dilute axion star. In principle, because of
quantum
fluctuations, the metastable BEC can decay towards a more stable
state -- a dense axion star if we take into account the repulsive
$\varphi^6$ term in the
self-interaction potential -- or collapse. In this section, we compute the
tunneling rate of the BEC and the lifetime of the metastable state by using the
instanton theory (a pedagogical exposition of this theory is
presented in \cite{tunnellong}). This path integral formulation lends itself
naturally to the
study of the semiclassical limit $\hbar\rightarrow 0$ via a steepest-descent
approach. As explained in
the Introduction, we use a Gaussian ansatz
and reduce the problem to the tunneling rate of a particle in a one
dimensional potential following the approach of Stoof \cite{stoof}.

\subsection{General expression}

\begin{figure}
\begin{center}
\includegraphics[clip,scale=0.3]{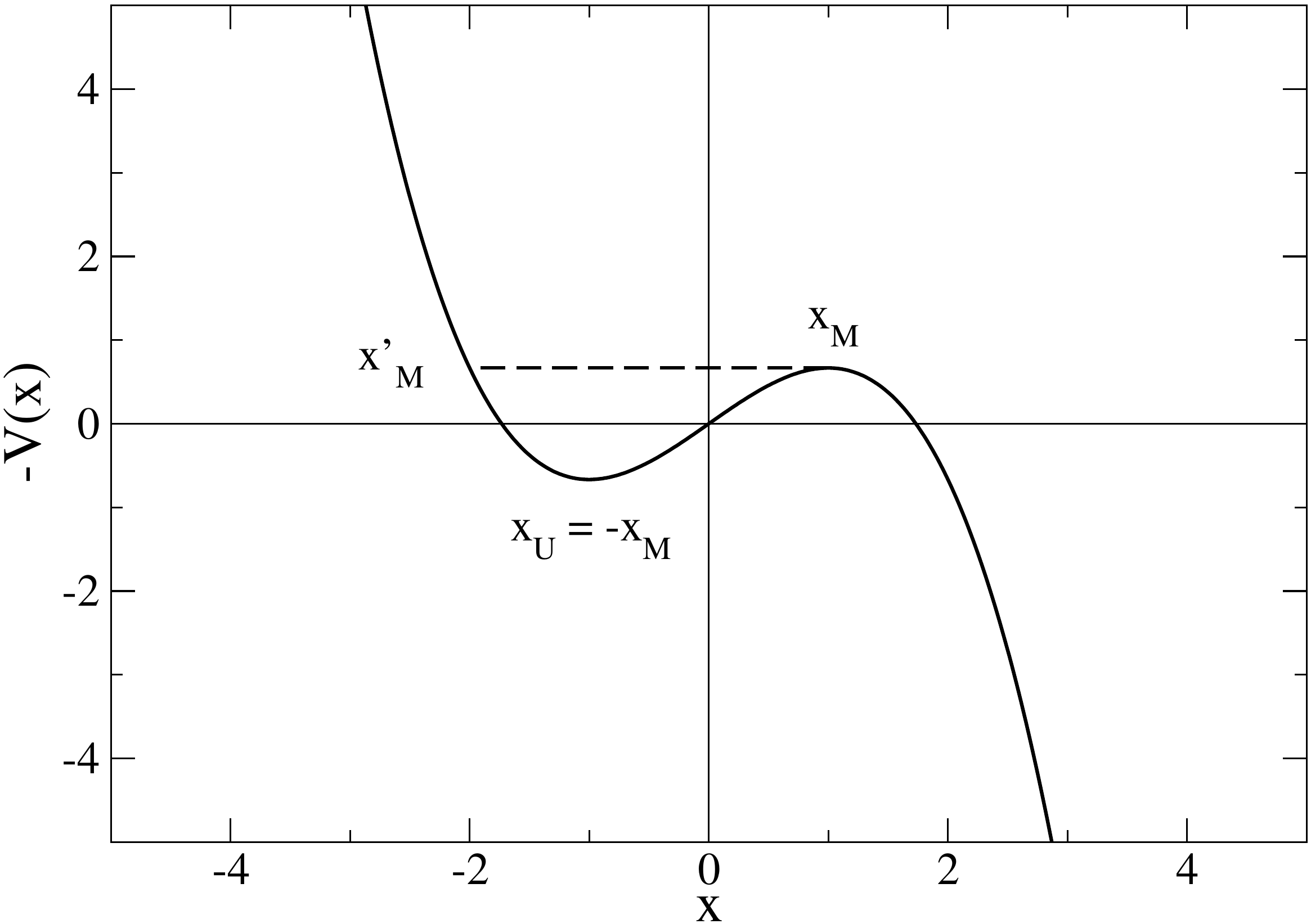}
\caption{Inverted potential occuring in the instanton theory (the dashed line
locates the bounce).}
\label{bounce}
\end{center}
\end{figure}

The equation of motion of
the fictive particle representing the BEC is given by Eq. (\ref{gte3}).
Classically ($\hbar=0$), the fictive particle can be in equilibrium in the
local minimum $R_M$ of the potential $V(R)$. If slightly displaced from its
equilibrium position, it will oscillate with a
pulsation $\omega_M^2=V''(R_M)/\alpha M$. However, because of
quantum fluctuations ($\hbar\neq 0$), this equilibrium state is metastable and
the particle can
cross the
potential barrier and escape. In the present formalism, quantum fluctuations
are incorporated in the Schr\"odinger equation 
\begin{eqnarray}
\label{qtr0}
i\hbar\frac{\partial\psi}{\partial t}=-\frac{\hbar^2}{2\alpha
M}\frac{d^2\psi}{dR^2}+V(R)\psi
\end{eqnarray}
for the fictive particle. In the
semiclassical limit $\hbar\rightarrow 0$,
the
quantum tunneling rate of the BEC 
is given by 
\begin{eqnarray}
\label{qtr1}
\Gamma\sim A\, e^{-B/\hbar},
\end{eqnarray}
where the prefactor $A$ is specificed below and the exponent $B$ is equal to
\begin{eqnarray}
\label{qtr2}
B=S[R_{b}(t)]-S[R_M],
\end{eqnarray}
where
\begin{eqnarray}
\label{qtr3}
S[R(t)]=\int \left\lbrack \frac{1}{2}\alpha M \left (\frac{dR}{dt}\right
)^2+V(R)\right\rbrack \, {d} t
\end{eqnarray}
is the euclidean action of the fictive particle representing the
BEC. It is obtained from the classical action by replacing $V(R)$ by
$-V(R)$ (this is achieved by making the Wick rotation $t\rightarrow -it$ in
the Feynman path integral \cite{tunnellong}). The
trajectory $R_b(t)$ occuring in $B$ is the one that makes
the euclidean action (\ref{qtr3}) extremal. This is the so-called instanton
(or bounce)
solution. Therefore, the bounce exponent $B$ is equal to the
value of the euclidean
(imaginary-time) action evaluated along the bounce trajectory (instanton).
The condition $\delta S=0$ leads to the equation 
\begin{equation}
\label{qtr4}
\alpha M\frac{{d}^2R_b}{{d}t^2}=V'(R_b),
\end{equation}
which is analogous to the classical equation of motion of a fictive
particle in the reversed potential $-V(R)$. If we
consider the classical equation of motion (\ref{gte3}) of the particle in
the potential
$V(R)$, the only solution consistent with the initial condition  $\dot R=0$ at
$R=R_M$ is $R(t)=R_M$ (see Fig. \ref{normalform}). It corresponds to the stable
equilibrium state of the
original problem. When we make the Wick rotation, we are led to the equation of
motion (\ref{qtr4}) for the particle in the reversed 
potential $-V(R)$. There are now two solutions consistent with the initial
condition $\dot
R=0$ at $R=R_M$. The first solution is the trivial solution
$R(t)=R_M$ mentioned previously. The second solution is a nontrivial topological
solution which extends far from $R_M$. This is
the
standard example of an instanton. It starts at $t\rightarrow
-\infty$ from the top
of the hill  $R_M$ with zero initial velocity, rolls down the
hill, bounces off the wall at
the turning point $R'_M$ such that $V(R'_M)=V(R_M)$ at some
time $t_c$ (this defines the center of the instanton) and
returns to the top of the hill $R_M$ with zero velocity at $t\rightarrow
+\infty$ (see Fig. \ref{bounce}). Using the
classical analogy, the so-called ``bounce''
solution $R_b(t)$
has the property that the particle spends a very long time around $R_M$ but in
a relatively short time oscillates once in the potential minimum of
$-V(R)$. The
first integral of motion of Eq. (\ref{qtr4}) is
\begin{equation}
\label{qtr5}
E=\frac{1}{2}\alpha M{\dot R_b}^2-V(R_b),
\end{equation}
where $E$ is a constant that can be called the energy of the instanton. It is
determined by the initial condition $\dot R_b=0$ at $R_b=R_M$ giving
$E=-V(R_M)$. As a result, the equation of the instanton is
\begin{equation}
\label{qtr6}
\frac{1}{2}\alpha M{\dot R_b}^2=V(R_b)-V(R_M),
\end{equation}
or, equivalently,
\begin{equation}
\label{qtr7}
\dot R_{b}=\mp\sqrt{\frac{2}{\alpha M}\lbrack V(R_{b})-V(R_M)\rbrack},
\end{equation}
where we should use the sign $-$ before the bounce at $R'_M$ and the sign $+$
after the bounce. The  instanton profile is given by an
integral of the form 
\begin{equation}
\label{qtr8}
\int_{R'_M}^{R_{b}(t)} \frac{{d}R}{\sqrt{\lbrack
V(R)-V(R_M)\rbrack}}=\mp\sqrt{\frac{2}{\alpha M}}(t-t_c).
\end{equation}
An arbitrary parameter $t_c$ indicates its center (defined by $\dot
R_b(t_c)=0$).

It is now
easy to obtain a
closed expression
for the euclidean action of the instanton in the limit $t\rightarrow
+\infty$. Using Eq. (\ref{qtr6}) the bounce exponent
\begin{equation}
\label{qtr9}
B=\int_{-\infty}^{+\infty} \left\lbrack \frac{1}{2}\alpha M \left
(\frac{dR_b}{dt}\right
)^2+V(R_b)-V(R_M)\right\rbrack \, {d} t
\end{equation}
can be written under the equivalent forms 
\begin{eqnarray}
\label{qtr10}
B=\int_{-\infty}^{+\infty}  2\left\lbrack V(R_{b})-V(R_M)\right\rbrack \,
{d}t,
\end{eqnarray}
or
\begin{eqnarray}
\label{qtr11}
B
=\int_{-\infty}^{+\infty}  \alpha M{\dot R}_{b}^2\,
{d}t.
\end{eqnarray}
The last integral can be rewritten as
\begin{eqnarray}
\label{qtr12}
B=2\int_{-\infty}^{t_c}  \alpha M {\dot R}_{b}^2\,
{d}t=2\int_{R_M}^{R'_M}  \alpha M {\dot R}_{b}\,
{d}R_{b},
\end{eqnarray}
leading to [see Eq. (\ref{qtr7})]
\begin{eqnarray}
\label{qtr13}
B=2\int_{R'_M}^{R_M}  \sqrt{2\alpha M\lbrack V(R)-V(R_M)\rbrack}\,
{d}R,
\end{eqnarray}
where we recall that $R'_M$ is the turning point (bounce) defined by the
condition
$V(R'_M)=V(R_M)$. If we use  the first two expressions to compute $B$, we have
to explicitly determine the trajectory of the instanton (bounce). If we use
the third expression, this is not necessary. We just need to know the
expression of the potential $V(R)$. This leads to the following expression of
the  quantum tunneling rate
\begin{equation}
\label{qtr14}
\Gamma\sim A\,
e^{-\frac{2}{\hbar}\int_{R'_M}^{R_M} \sqrt{2\alpha M\lbrack
V(R)-V(R_M)\rbrack}\,
{d}R}.
\end{equation}
This expression, which is valid in the semi-classical approximation
$\hbar\rightarrow
0$, can also be obtained by using the WKB method to find the
transmission
amplitude across the potential barrier \cite{llquantique}. Therefore, it is
oftentimes
called the WKB transmittivity formula. 

In many applications, the exponential
behavior of the tunneling rate is sufficient. The calculation
of the prefactor $A$ is more involved. It requires the determination of a
fluctuation determinant which was obtained by Duru {\it et
al.} \cite{duru} using the method of Gel'fand and Yaglom
\cite{gy}. This leads to the following expression of the prefactor
\begin{eqnarray}
\label{qtr15}
A=\sqrt{\frac{\alpha M\omega_Mv_M^2}{\pi\hbar}},
\end{eqnarray}
where $v_M$, which depends on the details of the potential, is determined by
the asymptotic behavior of the instanton solution via the formula
\begin{equation}
\label{qtr16}
R_{b}(t)\simeq {R_M}-\frac{v_M}{\omega_M}e^{-\omega_M|t|} \qquad (t\rightarrow
\pm\infty).
\end{equation}
The complete expression of the tunneling rate including the prefactor is
therefore
\begin{eqnarray}
\label{qtr17}
\Gamma\sim \sqrt{\frac{\alpha
M\omega_Mv_M^2}{\pi\hbar}}e^{-\frac{2}{\hbar}\int_{R'_M}^{R_M} \sqrt{2\alpha
M\lbrack V(R)-V(R_M)\rbrack}\,
{d}R}.
\end{eqnarray}
Finally, the typical lifetime of the
metastable state can be estimated by 
\begin{equation}
\label{qtr17b}
t_{\rm life}\sim \Gamma^{-1}
\sim\sqrt{\frac{\pi\hbar}{\alpha M\omega_Mv_M^2}}
e^{\frac{2}{\hbar}\int_{R'_M}^{R_M} \sqrt{2\alpha M\lbrack V(R)-V(R_M)\rbrack}\,
{d}R}.
\end{equation}

\subsection{Expression valid close to the maximum mass}

In this section, we determine the quantum tunneling rate of the BEC  close to
the maximum mass $M_{\rm max}$ by using the normal form of the
potential close to a saddle-center bifurcation
given by Eq. (\ref{nfcmm1}). It is
convenient to set $x=R-R_*$. In that
case, the potential  can be rewritten as 
\begin{eqnarray}
\label{qtr18}
V(x)=\frac{1}{3}ax^3-bx,
\end{eqnarray}
where $a$ and $b$ are two positive constants given by
\begin{eqnarray}
a=\frac{V_0}{R_*^3}\quad {\rm and} \quad b=\frac{2V_0}{R_*}\left
(1-\frac{M}{M_{\rm
max}}\right ).
\label{qtr19}
\end{eqnarray}
For simplicity, we have taken the additional constant in the potential equal to
zero (this is possible without restriction of generality since only differences
of potential occur in our problem). The potential (\ref{qtr18})
presents a local minimum and a local maximum (see Fig. \ref{normalform}). The
local minimum of $V(x)$ is located at $x_M=\sqrt{b/a}$
and the value of the potential at that point is
$V(x_M)=-(2/3)ax_M^{3}$. The maximum of $V(x)$ is located at $x_U=-x_M$ and
the value of the potential at that point is $V(x_U)=(2/3)ax_M^{3}$. The
bouncing (or escape) point $x'_M$ where
$V(x'_M)=V(x_M)$ is given by $x'_M=-2x_M$. Finally,
we note that $V(x)=0$ for $x=0$ and for $x=\pm\sqrt{3b/a}$.
With these notations, the potential (\ref{qtr18}) can be
rewritten as
\begin{equation}
\label{qtr20}
V(x)-V(x_M)=a\left (\frac{1}{3}x^3-x_M^2x+\frac{2}{3}x_M^3\right ).
\end{equation}
The roots of the third degree equation defined by the term in parenthesis in
Eq. (\ref{qtr20}) are
$x_M$ (double root) and $x'_M$ (single root). We then find that the
potential (\ref{qtr18}) can be written as
\begin{equation}
\label{qtr22}
V(x)-V(x_M)=\frac{a}{3}(x-x_M)^2(x-x'_M).
\end{equation}
For future use, we note that the barrier of potential $\Delta V=V(x_U)-V(x_M)$
is
\begin{eqnarray}
\label{attr6}
\Delta V=\frac{4}{3} a x_M^3.
\end{eqnarray} 
On the other hand, the square pulsations ($\omega^2=V''(x)/\alpha M$) of the
fictive particle at the metastable at unstable positions are
\begin{eqnarray}
\label{bttr6}
\omega_M^2=\frac{2}{\alpha M}\sqrt{ab}\qquad {\rm and}\qquad
\omega_U^2=-\frac{2}{\alpha M}\sqrt{ab}.
\end{eqnarray}

When $\hbar\rightarrow 0$, the quantum tunneling rate of
the BEC is given by Eq. (\ref{qtr17}). We propose
two methods to compute the bounce exponent $B$ in the exponential
factor using respectively the WKB
formula and the instanton solution. We also compute the prefactor $A$ of the
tunneling rate.

\subsubsection{The WKB formula}

The expression of
$B$ can be
obtained from the WKB formula (\ref{qtr13}). When the
potential is given by Eq. (\ref{qtr22}), the integral appearing in Eq.
(\ref{qtr13})
takes the explicit form
\begin{equation}
\label{q22}
B=2\sqrt{\frac{2\alpha M a}{3}}\int_{-2x_M}^{x_M} 
(x_M-x)\sqrt{x+2x_M}\,
{d}x.
\end{equation}
With the change of variables
\begin{equation}
\label{q23}
X=\sqrt{\frac{x}{3x_M}+\frac{2}{3}},
\end{equation}
it can be rewritten as
\begin{equation}
\label{q24}
B=36\sqrt{2\alpha M a}x_M^{5/2}\int_{0}^{1}
(1-X^2)X^2\,
{d}X.
\end{equation}
Using the identity
\begin{equation}
\label{q25}
\int_{0}^{1} 
(1-X^2)X^2\,
{d}X=\frac{2}{15},
\end{equation}
we obtain
\begin{equation}
\label{q26}
B=\frac{24}{5}\sqrt{2\alpha M a}x_M^{5/2}.
\end{equation}

\subsubsection{The instanton solution}

\begin{figure}
\begin{center}
\includegraphics[clip,scale=0.3]{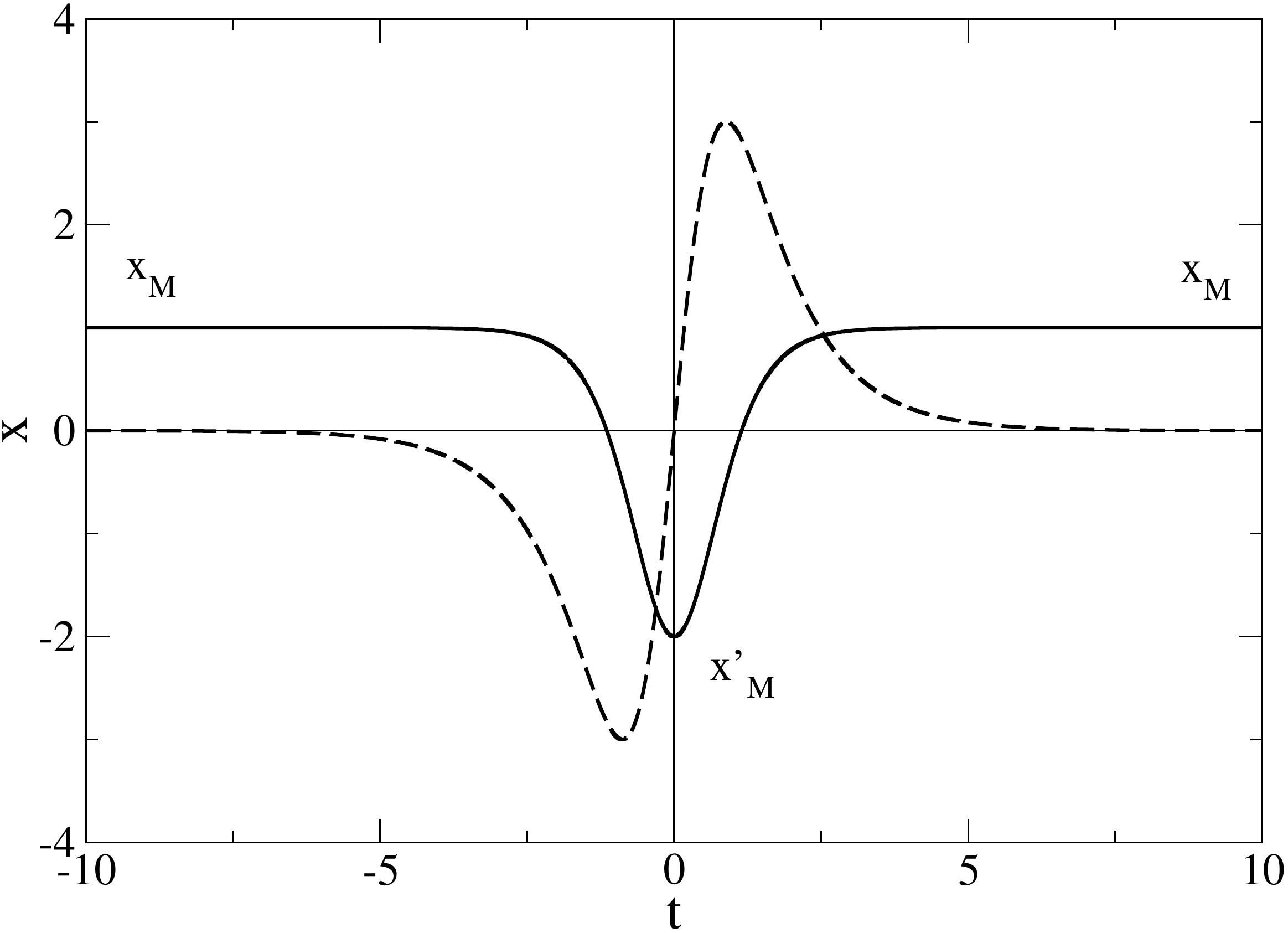}
\caption{Quantum instanton (bounce) $x_b(t)$ close to the maximum mass
(the dashed line corresponds to the velocity ${\dot x}_b(t)$).}
\label{instantonQ}
\end{center}
\end{figure}

The expression of
$B$ can also be
obtained from Eqs. (\ref{qtr10}) and
(\ref{qtr11}) by explicitly calculating the instanton solution. The
instanton (bounce) is determined by Eq. (\ref{qtr8}). When the potential
is
given by Eq. (\ref{qtr22}), this equation becomes
\begin{equation}
\label{q28}
\int^{x_{b}(t)} \frac{{d}x}{(x_M-x)\sqrt{x+2x_M}}=-\sqrt{\frac{2a}{3\alpha M}}t.
\end{equation}
With the change of variables
\begin{equation}
\label{q29}
X=\sqrt{\frac{x}{3x_M}+\frac{2}{3}},
\end{equation}
it can be rewritten as 
\begin{equation}
\label{q30}
\int^{\sqrt{\frac{x_{b}(t)}{3x_M}+\frac{2}{3}}}
\frac{{d}X}{1-X^2}=-\sqrt{\frac{ax_M}{2\alpha M}}t.
\end{equation}
Using the identity 
\begin{equation}
\label{q31}
\int\frac{{d}X}{1-X^2}=\tanh^{-1}(X)\qquad (-1<X<1),
\end{equation}
we  obtain
\begin{eqnarray}
\label{q32}
x_{b}(t)&=&x_M \left\lbrack 3\tanh^2\left
(\sqrt{\frac{ax_M}{2\alpha M}}t\right )-2\right\rbrack\nonumber\\
&=&{x_M}\left
\lbrack
1-\frac{3}{\cosh^2\left
(\sqrt{\frac{a x_M}{2\alpha M}}t\right )}\right\rbrack.
\end{eqnarray}
This is the instanton (bounce) solution (see Fig.
\ref{instantonQ}). We have
chosen
the origin of time so that the
instanton center is at
$x'_M=-2x_M$ (bouncing point) at
$t=0$. For
$t\rightarrow \pm\infty$, we have $x_b(t)\rightarrow x_M$. It is
precisely this type of solutions, which
approach a
static limit in the distant past and future, that are refered to as
``instantons''. The
velocity of the fictive particle associated with the instanton solution is
\begin{equation}
\label{q33}
{\dot x}_{b}(t)=6{x_M}\sqrt{\frac{ax_M}{2\alpha M}} \frac{\sinh\left
(\sqrt{\frac{a x_M}{2\alpha M}}t\right )}{\cosh^3\left
(\sqrt{\frac{a x_M}{2\alpha M}}t\right )}.
\end{equation}
Substituting this expression into Eq. (\ref{qtr11}) we obtain
\begin{equation}
\label{q34}
B=18\sqrt{2\alpha
M a}x_M^{5/2}\int_{-\infty}^{+\infty}\frac{\sinh^2(x)}{\cosh^6(x)
}\,
{d}x.
\end{equation}
Using the identity 
\begin{equation}
\label{q35}
\int_{-\infty}^{+\infty}\frac{\sinh^2(x)}{\cosh^6(x)}\,
{d}x=\frac{4}{15},
\end{equation}
we recover Eq. (\ref{q26}). The same expression can also be obtained from Eq.
(\ref{qtr10}).

\subsubsection{The prefactor}

To obtain the prefactor of the tunneling rate given by Eq. (\ref{qtr15}) we
first note that $\omega_M^2=V''(x_M)/\alpha M=2 a x_M/\alpha M$. On the
other hand, from Eq. (\ref{q32}),
we have
\begin{equation}
\label{q36}
x_{b}(t)\simeq x_M-12x_Me^{-\sqrt{\frac{2ax_M}{\alpha M}}|t|} \qquad 
(t\rightarrow
\pm\infty).
\end{equation}
Comparing this asymptotic behavior with the expression from Eq.
(\ref{qtr16}), we
obtain
\begin{equation}
\label{q37}
v_M=12\left (\frac{2a}{\alpha M}\right )^{1/2}x_M^{3/2}.
\end{equation}
Therefore, the prefactor of the tunneling
rate is 
\begin{equation}
\label{q39}
A=12\left (\frac{8a^3}{\pi^2 \alpha M \hbar^2}\right )^{1/4}x_M^{7/4}.
\end{equation}
Combining Eqs.
(\ref{qtr1}), (\ref{q26}) and (\ref{q39}), we find that the
complete expression
of the quantum tunneling rate of the BEC close to the maximum mass is given by
\begin{equation}
\label{q40}
\Gamma\sim 12\left (\frac{8a^3}{\pi^2 \alpha M \hbar^2}\right
)^{1/4}x_M^{7/4}e^{-\frac{24}{5\hbar}\sqrt{2\alpha M a}x_M^{5/2}}.
\end{equation}
Returning to the original variables, we get
\begin{eqnarray}
\label{q41}
\Gamma\sim 12\left (\frac{8}{\pi^2}\right )^{1/4}\left\lbrack
2\left (1-\frac{M}{M_{\rm max}}\right
)\right\rbrack^{7/8}(\alpha\sigma)^{1/4}\sqrt{N}\nonumber\\
\times e^{-\frac{24}{5}\sqrt{2}\left\lbrack
2\left (1-\frac{M}{M_{\rm max}}\right
)\right\rbrack^{5/4}\sqrt{\alpha\sigma}N}t_D^{-1},
\end{eqnarray}
where we have introduced the particle number $N=M/m$ and we
recall that the
above expression is valid for  $M\rightarrow M_{\rm max}$. We note that the
bounce exponent scales as $B\propto \left
(1-{M}/{M_{\rm max}}\right
)^{5/4}$ and the prefactor as $A\propto \left (1-{M}/{M_{\rm
max}}\right
)^{7/8}$. These are the same scalings as those obtained in
Refs. \cite{leggett,huepe} for nongravitational BECs. These scalings are
universal since they just depend on the normal form of the potential close to a
saddle-center bifurcation.

\section{Thermal tunneling rate of the BEC}
\label{sec_ttr}

In addition to quantum fluctuations, the BEC may also experience thermal
fluctuations that can destabilize the metastable equilibrium state. Indeed,
because of thermal fluctuations the system can overcome the energy barrier
between the metastable state and the unstable state and collapse. We provide
here a very heuristic treatment of thermal fluctuations in a BEC, using an
analogy with the Kramers \cite{kramers} problem in Brownian theory (a similar
approach has been used in \cite{stoof,huepe} for nongravitational BECs and in
\cite{lifetime} for
globular clusters).

In Sec. \ref{sec_ga}, making a Gaussian ansatz, we have reduced
the
original problem (solving the
GPP equations (\ref{gpp1}) and (\ref{gpp2})) to the simpler mechanical problem
of a particle with mass $\alpha
M$ in a potential $V(R)$ governed by the deterministic equation
(\ref{gte3}). Within this framework, we have taken into
account quantum
fluctuations  in Sec. \ref{sec_qtr} by replacing the deterministic equation
(\ref{gte3}) by the
Schr\"odinger
equation (\ref{qtr0}). Similarly, we
can take
thermal fluctuations into
account by replacing
the determinsitic equation (\ref{gte3}) by a  stochastic Langevin equation of
the form
\begin{eqnarray}
\label{ttr0}
\alpha M\frac{d^2R}{dt^2}+\xi\alpha
M\frac{dR}{dt}=-\frac{dV}{dR}+\sqrt{2\xi\alpha M k_B T}\, \eta(t),
\end{eqnarray} 
where $\eta(t)$ is a Gaussian white noise with $\langle \eta(t)\rangle=0$ and
$\langle \eta(t)\eta(t')\rangle=\delta (t-t')$. This equation involves a
friction force
characterized by a friction coefficient $\xi$ and a random force
whose strength
is measured by the temperature $T$.  These two effects
arise simultaneously on account of the fluctuation-dissipation theorem
encapsulated in the Einstein relation $D=\xi
k_B T/\alpha M$, where $D$ is the diffusion coefficient. The
thermal tunneling (or thermal activation) rate is of the general form
\begin{eqnarray}
\label{ttr1}
\Gamma\sim A\, e^{-\Delta V/k_B T},
\end{eqnarray} 
where $\Delta V=V(R_U)-V(R_M)$ is the potential barrier between the metastable
state and the unstable state and $A$ is a prefactor discussed below. The
expression (\ref{ttr1}) is valid when $k_B T\ll
\Delta V$. The exponential term in Eq. (\ref{ttr1}) was obtained long ago by
Arrhenius \cite{arrhenius} from an empirical analysis of chemical reaction
rates
and is called the Arrhenius law. It was later justified by Kramers
\cite{kramers} from the detailed study of the stochastic motion of a Brownian
particle past a potential barrier. The prefactor has different expressions
depending on the considered regime.  Kramers \cite{kramers} obtained the general
formula 
\begin{eqnarray}
\label{ttr2}
\Gamma\sim \frac{\omega_M}{2\pi |\omega_U|}\left\lbrack
\sqrt{\frac{\xi^2}{4}+|\omega_U|^2}-\frac{\xi}{2}\right\rbrack \, e^{-\Delta
V/k_B T},
\end{eqnarray} 
where we recall that $\omega_M^2=V''(R_M)/\alpha M>0$ and
$\omega_U^2=V''(R_U)/\alpha M<0$. This formula was derived from a Fokker-Planck
equation in phase space (Kramers equation). In the strong
friction
limit $\xi\rightarrow +\infty$, Eq. (\ref{ttr2}) reduces to 
\begin{eqnarray}
\label{ttr3}
\Gamma\sim \frac{\omega_M |\omega_U|}{2\pi\xi}\, e^{-\Delta V/k_B T}.
\end{eqnarray} 
This asymptotic result can be directly obtained from a Fokker-Planck
equation in position space (Smoluchowski equation). In the  weak
friction limit $\xi\rightarrow 0$, Eq. (\ref{ttr2}) reduces to 
\begin{eqnarray}
\label{ttr4}
\Gamma\sim \frac{\omega_M}{2\pi}\, e^{-\Delta V/k_B T},
\end{eqnarray}
which corresponds to the result of the transition state theory.
However, using a more careful treatment, Kramers \cite{kramers}  showed that
this
aymptotic formula is not perfectly correct and that it must be replaced by
\begin{eqnarray}
\label{ttr5}
\Gamma\sim \frac{\xi I_U}{k_B T} \frac{\omega_M}{2\pi}\, e^{-\Delta V/k_B T},
\end{eqnarray} 
where $I_U\sim 2\pi \Delta V/\omega_M$ is the action of the path at the barrier
peak. This more accurate
expression shows that, when $\xi\rightarrow 0$, the thermal tunneling rate
$\Gamma$ vanishes proportionally to $\xi$ instead of tending to a constant. For
sufficiently large values of $\xi$, the
expressions (\ref{ttr2}) and (\ref{ttr3}) become valid.

Close to the maximum mass, using the normal form of the potential
(\ref{qtr22}), we find that the thermal tunneling rate of the BEC based on the
Arrhenius law (\ref{ttr1}) is given by
\begin{eqnarray}
\label{ttr6}
\Gamma\propto 
e^{-\frac{4}{3}\left\lbrack
2\left
(1-\frac{M}{M_{\rm max}}\right )\right\rbrack^{3/2}\nu\eta N},
\end{eqnarray}
where we have introduced the
particle number
$N=M/m$ and the dimensionless inverse temperature
\begin{eqnarray}
\label{ttr7}
\eta=\frac{GM_{\rm max} m}{R_* k_B T}.
\end{eqnarray}
We recall that Eq. (\ref{ttr6}) is valid for $M\rightarrow M_{\rm max}$.
The
complete expression of the thermal tunneling rate based on the Kramers formula
(\ref{ttr2}) is 
\begin{eqnarray}
\label{ttr8}
\Gamma\sim \frac{1}{2\pi}\left\lbrack
\sqrt{\frac{\xi^2}{4}+\frac{2}{t_D^2}\left\lbrack
2\left (1-\frac{M}{M_{\rm max}}\right
)\right\rbrack^{1/2}}-\frac{\xi}{2}\right\rbrack \nonumber\\
\times e^{-\frac{4}{3}\left\lbrack
2\left
(1-\frac{M}{M_{\rm max}}\right )\right\rbrack^{3/2}\nu\eta N}.
\end{eqnarray} 
In the strong friction limit $\xi\rightarrow +\infty$, we get
\begin{equation}
\label{ttr9}
\Gamma\sim \frac{1}{\pi\xi t_D^2} \left\lbrack 2\left
(1-\frac{M}{M_{\rm max}}\right )\right\rbrack^{1/2}\,
e^{-\frac{4}{3}\left\lbrack
2\left
(1-\frac{M}{M_{\rm max}}\right )\right\rbrack^{3/2}\nu\eta N}.
\end{equation}
In the weak friction limit $\xi\rightarrow 0$, we obtain
\begin{equation}
\label{ttr10}
\Gamma\sim \frac{1}{\sqrt{2}\pi t_D}\left\lbrack
2\left (1-\frac{M}{M_{\rm max}}\right
)\right\rbrack^{1/4}\, e^{-\frac{4}{3}\left\lbrack
2\left
(1-\frac{M}{M_{\rm max}}\right )\right\rbrack^{3/2}\nu\eta N},
\end{equation}
although this formula is not fully correct as mentioned above. We note that the
potential barrier scales as $\Delta V\propto
\left
(1-{M}/{M_{\rm max}}\right
)^{3/2}$. This is the same
scaling as the one obtained in \cite{huepe} for nongravitational BECs and in
\cite{lifetime} for globular
clusters. This scaling is
universal since it just depends on the normal form of the potential close to a
saddle-center bifurcation.

{\it Remark:} The thermal tunneling rate of the BEC can
also be obtained by applying the instanton theory to the generalized stochastic
GPP and quantum Smoluchowski-Poisson equations \cite{tunnellong} (see
also Appendix \ref{sec_sgl} for the related
stochastic Ginzburg-Landau-Poisson equations).

\section{Correction of the critical mass due to quantum and thermal 
fluctuations}
\label{sec_k}

Let us summarize the preceding results. A self-gravitating BEC with
an attractive self-interaction can exist only
below a maximum mass $M_{\rm max}$ \cite{prd1,prd2}. For $M<M_{\rm max}$
and $R>R_*$ the BEC is in a metastable state (local but not global minimum of
energy) corresponding to a dilute axion star. Because of quantum fluctuations it
can decay into a more
stable state (dense axion star) if we account for the repulsive
self-interaction
between the bosons, or collapse if there is no repulsive
self-interaction. The lifetime of the metastable
state due to
quantum fluctuations can be estimated by $t_{\rm life}^{\rm Q}\sim
1/\Gamma_{\rm Q}$ where $\Gamma_{\rm Q}$ is the quantum tunneling rate of the
BEC. According to
Eq. (\ref{q41}), we have 
\begin{eqnarray}
\label{k1}
t_{\rm life}^{\rm Q}\sim \frac{1}{12}\left (\frac{\pi^2}{8}\right
)^{1/4}\left\lbrack
2\left (1-\frac{M}{M_{\rm max}}\right
)\right\rbrack^{-7/8}\frac{1}{(\alpha\sigma)^{1/4}}\frac{1}{\sqrt{N}}\nonumber\\
\times e^{\frac{24}{5}\sqrt{2}\left\lbrack
2\left (1-\frac{M}{M_{\rm max}}\right
)\right\rbrack^{5/4}\sqrt{\alpha\sigma}N}t_D.\qquad
\end{eqnarray}
The quantum lifetime of dilute axion stars scales as 
\begin{eqnarray}
\label{k2}
t_{\rm life}^{\rm Q}\sim e^{N}\, t_D,
\end{eqnarray}
except close
to the critical point. Since $N$ is very large ($N=7.21\times 10^{56}$ for QCD
axions
and $N=5.09\times 10^{95}$ for ULAs), the
lifetime of
a metastable state is considerable \cite{phi6}. As a matter of fact, metastable
states can
be considered as stable states. Only extraordinarily close to the maximum mass
will their
lifetime decrease. In principle, the BEC will collapse at a mass  
$M_{\rm crit}$ smaller than $M_{\rm max}$. The mass
at which the BEC collapses because of quantum tunneling can
be estimated by writing that the exponent of the exponential term in Eq.
(\ref{k1}) is of order unity.
This gives
\begin{equation}
\label{k3}
M_{\rm crit}^{\rm Q}\sim M_{\rm max}\left (1-0.103\, N^{-4/5}\right
).
\end{equation}
It displays the scaling $N^{-4/5}$. For large values of $N$, which is the case
for axion stars, this correction is
extremely
small and can be neglected. Therefore, the value of the maximum mass of axion
stars \cite{prd1,prd2} is essentially unaffected by quantum tunneling. 
However,
corrections due to quantum tunneling could be observed in laboratory
experiments and numerical simulations attempting to mimic ``axion stars''
because, in that case, the number of particles will be necessarily reduced as
compared to reality.

Similar results are obtained if we account for thermal tunneling. The lifetime
of the metastable
state due to thermal fluctuations can be estimated by $t_{\rm life}^{\rm T}\sim
1/\Gamma_{\rm T}$ where $\Gamma_{\rm T}$ is the thermal tunneling rate of the
BEC. According to
Eq. (\ref{ttr6}), we have
\begin{eqnarray}
\label{k1b}
t_{\rm life}^{\rm T}\propto 
e^{\frac{4}{3}\left\lbrack
2\left
(1-\frac{M}{M_{\rm max}}\right )\right\rbrack^{3/2}\nu\eta N},
\end{eqnarray}
where the value of the prefactor is discussed in Sec. \ref{sec_ttr}. The
thermal lifetime of dilute axion stars scales as 
\begin{eqnarray}
\label{k4}
t_{\rm life}^{\rm T}\propto e^{N},
\end{eqnarray}
except close
to the critical point. The mass
at which the BEC collapses because of thermal tunneling can
be estimated by 
\begin{equation}
\label{k5}
M_c^{\rm T}\sim M_{\rm max}\left (1-0.762\, N^{-2/3}\right
),
\end{equation}
where we have taken $\eta\sim 1$ for simplicity. It displays the scaling
$N^{-2/3}$. The same scaling was found in \cite{katzokamoto,lifetime} in
the case of globular clusters. 
For axion stars, this correction is extremely
small and can be neglected. Therefore, the value of the maximum
mas of axion stars \cite{prd1,prd2} is not altered by thermal effects.

\section{Finite size scaling close to the maximum mass}
\label{sec_size}

In Ref. \cite{bectcoll} we have studied the collapse time of dilute axion stars
when $M>M_{\rm max}$. For $M\rightarrow M_{\rm
max}^+$, we have found that
\begin{equation}
\label{size1}
\frac{t_{\rm coll}}{t_D}\sim \frac{2.90}{(M/M_{\rm max}-1)^{1/4}}.
\end{equation}
This study neglects quantum and thermal fluctuations. As a result, it
predicts that the collapse time is infinite
when $M=M_{\rm max}$. Actually, because of 
 quantum and thermal fluctuations, the collapse time should be
large but finite at $M=M_{\rm max}$. We develop below an argument to estimate
the finite size scaling of the collapse time close to
the
maximum mass and its finite value at $M=M_{\rm max}$.

For $M<M_{\rm max}$, we have found that the quantum lifetime of dilute axion
stars is given by Eq. (\ref{k1}). Finite size effects enter in the expression of
the metastable state lifetime in the combination $N(1-M/M_{\rm max})^{5/4}$. If
we assume that a similar combination enters in the expression of the collapse
time for $M>M_{\rm max}$ we expect a scaling of the form
\begin{equation}
\label{size2}
\frac{t_{\rm coll}}{t_D}\propto (M/M_{\rm max}-1)^{-1/4} F\left\lbrack
N(M/M_{\rm max}-1)^{5/4}\right\rbrack
\end{equation}
with $F(x)\rightarrow 1$ for $x\rightarrow +\infty$ in order to recover Eq.
(\ref{size1}) when $N\rightarrow +\infty$. At $M=M_{\rm max}$, the singular
factor $(M/M_{\rm max}-1)^{-1/4}$ must cancel out implying that $F(x)\sim
x^{1/5}$ for $x\rightarrow 0$. Therefore, at $M=M_{\rm max}$, the collapse time
taking into account quantum fluctuations scales as
 \begin{equation}
\label{size3}
t_{\rm coll}^{\rm Q}\propto N^{1/5}\, t_D, \qquad (M=M_{\rm max}).
\end{equation}

We can make similar calculations to account for thermal fluctuations. In that
case, finite size effects enter in the expression of the
metastable state lifetime in the combination $N(1-M/M_{\rm max})^{3/2}$ [see
Eq. (\ref{k1b})]. We therefore expect a scaling of the form
\begin{equation}
\label{size4}
\frac{t_{\rm coll}}{t_D}\propto (M/M_{\rm max}-1)^{-1/4} F\left\lbrack
N(M/M_{\rm max}-1)^{3/2}\right\rbrack
\end{equation}
with $F(x)\rightarrow 1$ for $x\rightarrow +\infty$ and  $F(x)\rightarrow
x^{1/6}$ for
$x\rightarrow 0$. Therefore, at $M=M_{\rm max}$, the collapse time taking into
account thermal fluctuations scales as
 \begin{equation}
\label{size5}
t_{\rm coll}^{\rm T}\propto N^{1/6}\, t_D, \qquad (M=M_{\rm max}).
\end{equation}

For QCD axions with mass  $m=10^{-4}\,
{\rm eV}/c^2$ and self-interaction $a_s=-5.8\times
10^{-53}\, {\rm m}$, the maximum mass is  $M_{\rm max}^{\rm
exact}=6.46\times 10^{-14}\,
M_{\odot}$ and the minimum radius is $(R_{99}^*)^{\rm exact}=227\, {\rm
km}$. As a result, the typical number of axions 
is $N\sim 10^{57}$ and the typical dynamical time is $t_D\sim 10\, {\rm hrs}$.
Then, we get $t_{\rm coll}^Q\sim 10^8\, {\rm yrs}$ and $t_{\rm coll}^T\sim
10^6\,
{\rm yrs}$. The collapse time of QCD axion stars at criticality is smaller
than the age of the Universe ($\sim 14\times
10^{9}\, {\rm yrs}$). 

For ULAs with  mass 
$m=2.19\times 10^{-22}\, {\rm eV}/c^2$  and 
self-interaction $a_s=-1.11\times 10^{-62}\, {\rm fm}$, the
maximum mass is  $M_{\rm max}^{\rm
exact}=10^8\, M_{\odot}$ and the minimum radius is
$(R^*_{99})^{\rm
exact}=1\, {\rm
kpc}$. As a result, the typical number of axions
is $N\sim 10^{96}$ and the typical dynamical time is $t_D\sim 10^8\, {\rm yrs}$.
Then, we
get $t_{\rm coll}^Q\sim 10^{27}\, {\rm yrs}$ and $t_{\rm coll}^T\sim 10^{24}\,
{\rm yrs}$. The collapse time of axion stars (or of the quantum core of DM
halos) made of ULAs at criticality is much larger than the age of the Universe
($\sim 14\times
10^{9}\, {\rm yrs}$).

{\it Remark:} We note that the scalings from Eqs.
(\ref{size3}) and (\ref{size5}) for the collapse time at criticality are also
valid for nongravitational BECs with an attractive self-interaction in a
confining trap (see footnote 10). They do not seem to have been reported
previously in that context.

\section{Conclusion}
\label{sec_conclusion}

In this paper, we have
computed the quantum and thermal tunneling rates of
dilute axion stars close to the maximum mass $M_{\rm max}$ \cite{prd1,prd2}. In
the quantum case, we have shown
that the bounce exponent
vanishes as $(1-M/M_{\rm max})^{5/4}$ and the amplitude as $(1-M/M_{\rm
max})^{7/8}$. In the thermal case, we have shown that the energy barrier
vanishes as $(1-M/M_{\rm max})^{3/2}$. The same scalings were previously
obtained in the
case of nongravitational
BECs with attractive self-interaction in a harmonic trap close to the
maximum particle number \cite{leggett,huepe}. The  scaling
for the bounce exponent of the quantum tunneling rate was also obtained long ago
by \cite{htww} in the case of neutron stars close to the
Oppenheimer-Volkoff maximum mass and the scaling for the thermal tunneling
rate
was also obtained  by \cite{lifetime} in the case of globular
clusters close to
the point of gravitational collapse. These scalings
reflect 
the universal form of the potential close to a saddle-center bifurcation.
However, despite these attenuation factors, the lifetime of  dilute axion
stars generically scales as $e^N\, t_D$ as anticipated in \cite{phi6}.
In the case of axion stars, the
number of bosons is very large ($N\sim 10^{50}-10^{100}$) implying that the 
lifetime of dilute axion
stars is considerable. As a matter of fact, these metastable states can be
considerved as stable states \cite{phi6,ebybh} except extremely 
close to the critical point. Barrier penetration is a notoriously
slow process. Indeed, similar results regarding the
very long lifetime
of metastable states have been previously obtained in the case of
systems with
long-range interactions \cite{langerspin,art,cdkramers,bmf}, neutron stars
\cite{htww}, quantum field
theory in the early
universe
\cite{linde77,linde80,linde81,linde83,gw,cook,billoire,steinhardt81,witten,
hm,chh},\footnote{In the context of quantum field theory,
it was believed in the 1970-1980s that the early
Universe, by cooling below some critical temperature $T_0$, had undergone a
first order phase
transition   from a metastable symmetic state $\varphi=0$
(false vacuum) to a
stable symmetry-breaking Higgs state $\varphi=\sigma$  (true
vacuum) giving mass to the particles
\cite{voloshin,stone1,stone2,coleman,cc,cdl}. The tunneling was expected
to proceed through the
formation of bubbles  like in the
liquid-gas phase transition \cite{langer}. However, it was soon realized that
the nucleation of bubbles, thermal or quantum, was a very rare event and that
the tunneling probability was
extremely small.  As a result, the lifetime of the metastable vacuum state
$\varphi=0$ in
gauge theories is usually extremely large
\cite{linde77,linde80,linde81,linde83,gw,cook,billoire,steinhardt81,witten,
hm,chh}, much larger than the
age of the Universe. In practice, in these scenarios,
the Universe remains in the supercooled symmetric vacuum state $\varphi=0$,
leading to a phase of inflation during which the scale factor increases
exponentially with
time
\cite{guthinflation,linde82,gw,sato,sato2,cook,billoire,witten,
steinhardt81,hm,as,astw}. Therefore, the
transition from the symmetric state $\varphi=0$ to the asymmetric state
$\varphi=\sigma$ does not occur at $T_0$, where
the asymmetric
state becomes energetically favored, but at a much lower temperature
$\sim T_c$ at which the symmetric state $\varphi=0$ becomes unstable.
Now that the Higgs mass has been measured and found to be much larger than the
value required in the previous scenarios, it is rather believed that
the early Universe experienced a second order phase transition from a symmetric
phase $\varphi=0$ at high temperatures ($T>T_c$) to a symmetry-breaking phase
$\varphi=\sigma$ at low temperatures ($T<T_c$).}  laboratory BECs
\cite{stoof,leggett,huepe},
and globular
clusters \cite{katzokamoto,lifetime}.\footnote{The very long lifetime of
metastable states justifies
the notion of statistical
equilibrium for self-gravitating systems \cite{lifetime}. It
is well-known since the
works of Antonov \cite{antonov} and Lynden-Bell and Wood \cite{lbw} that no
equilibrium state for self-gravitating systems exists in a strict
sense, even if they are
confined within a box in order to prevent their
evaporation or if we use the King model to take into account
tidal effects. They can
always increase their entropy at fixed energy and mass by
forming a ``core-halo'' structure made of a binary star (containing a very
negative potential energy) surrounded by a hot halo (containing a very positive
kinetic energy). In this sense, there is no global maximum of entropy at fixed
energy and mass. The system is ultimately expected to collapse (gravothermal
catastrophe). However, there exist metastable equilibrium states (local but not
global maxima of entropy at fixed energy and mass) if the energy is not too low.
If the system is initially in a metastable state (which is the most natural
situation), it must cross a huge barrier of entropy to collapse. This is
achieved by forming a condensed structure (or a binary star) similar to a
``critical droplet'' in the physics of phase transitions and nucleation
problems. This requires nontrivial three-body or higher correlations. This is a
very rare event whose probability scales as
$e^{-N}$. Therefore the lifetime of the metastable state scales as $e^N$. This
is larger or comparable to the age of the Universe making metastable states
fully relevant. Therefore, in practice, metastable states can be considered as
stable equilibrium states.} In all these examples, quantum and
thermal tunneling are very
rare processes.

More precisely, we can draw the following
conclusions depending on the number of particles in the system:

(i) In laboratory BECs, the number of bosons in a quantum gas  is moderate
($N\sim 1000$) so that quantum and thermal tunnelings, even if they are small,
can be observed and measured \cite{leggett,huepe}.

(ii) In globular clusters, the number of stars is of the order of $N\sim 10^6$.
This number is large but not gigantic. In particular, finite $N$ effects
can advance
the onset of
gravitational collapse (gravothermal catastrophe)
\cite{monaghan,katzokamoto,lifetime}. The
critical density contrast taking
into
account the finite number of particles is ${\cal
R}_c=709\times {\rm exp}(-3.30N^{-1/3})$ giving ${\cal
R}_c=686$ for $N=10^6$ instead of ${\cal
R}_c=709$. This may explain why
observations
reveal that a greater number
of globular clusters than is normally believed may already be in an advanced
stage of core collapse. 

(iii) In axion stars, the number of bosons is of the order of
$N\sim 10^{50}-10^{100}$. This number is gigantic so
that quantum and thermal tunnelings are completely negligible and the onset of
gravitational collapse (or the value of the maximum mass) is not
altered \cite{phi6}. Only extraordinarily close to the maximum
mass does their lifetime decrease.

Our results show that the tunneling
rate of dilute axion stars is usually negligible. This strengthens the
validity of our former results \cite{prd1,bectcoll,phi6} where this effect
was neglected from the start. This is because, in axion
stars, the number of bosons $N$  is huge. However, we may imagine that, in a
close future, it
will be possible to make laboratory experiments and numerical simulations of
``axion stars'' or self-gravitating BECs with an attractive self-interaction. In
such experiments, and in the first
generation of numerical simulations, the number of bosons $N$ will be
relatively small and tunneling effects may be measurable. Our results will be
useful to interpret such experiments and numerical
simulations. The general
methods presented in our paper may also find applications in other situations of
physical interest, beyond axion stars, where the tunneling rate is larger.

On the other hand, the present study allowed us to take into account the effect
of
fluctuations during the collapse of axion stars which were ignored
in our previous work \cite{bectcoll}. In particular, at the maximum
mass $M=M_{\rm max}$, we found that the collapse time scales as
$t_{\rm coll}^{\rm Q}\propto N^{1/5}\, t_D$ in the
quantum case  and  $t_{\rm coll}^{\rm T}\propto
N^{1/6}\, t_D$ in the thermal case instead of being infinite as implied by Eq.
(\ref{size1}) which does not take into account
quantum and thermal fluctuations \cite{bectcoll}. We then found that the
collapse time 
is smaller than age of universe for QCD axion stars but larger for axion
stars (or for the quantum core of DM halos) made of ULAs.

\appendix

\section{From the KGE equations to the GPP equations}
\label{sec_kge}

In this Appendix, we show that the GPP equations (\ref{gpp1}) and (\ref{gpp2})
can be derived from
the KGE equations in the nonrelativistic limit $c\rightarrow +\infty$. For the
sake of
generality, we take into account the expansion of
the Universe (the static case is recovered for $a=1$). We
specifically consider the case of dilute axion stars and relate the scattering
length $a_s$
appearing
in
the GP equation (\ref{gpp1}) to the axion decay constant $f$. This Appendix
follows
Secs. II and III of \cite{phi6}.

We consider the relativistic quantum field theory of a real
SF $\varphi({\bf r},t)$ with an action
\begin{equation}
S=\int d^4x\, \sqrt{-g} \, {\cal L}
\label{kge0a}
\end{equation}
associated with the Lagrangian density
\begin{equation}
{\cal L}=
\frac{1}{2}g^{\mu\nu}\partial_\mu\varphi\partial_{\nu}
\varphi-\frac{m^2c^2}{2\hbar^2}\varphi^2-V(\varphi)+\frac {
c^4}{16\pi G}R,
\label{kge0b}
\end{equation}
where $g_{\mu\nu}$ is the metric tensor, $g$ is its determinant, $R$ is the
Ricci scalar, and $V(\varphi)$ is the potential of the SF. The least
action principle $\delta S=0$ leads to the KGE 
equations (see, e.g., \cite{abrilph}) 
\begin{equation}
\Box\varphi+\frac{m^2c^2}{\hbar^2}\varphi+\frac{dV}{d\varphi}=0,
\label{kge1}
\end{equation}
\begin{equation}
R_{\mu\nu}-\frac{1}{2}g_{\mu\nu}R=\frac{8\pi G}{c^4}T_{\mu\nu},
\label{kge2}
\end{equation}
where $\Box=D_{\mu}(g^{\mu\nu}\partial_{\nu})=\frac{1}{\sqrt{-g}}
\partial_\mu(\sqrt{-g}\, g^{\mu\nu}\partial_\nu)$ is the d'Alembertian
in a curved spacetime and
\begin{equation}
T_{\mu\nu}=\partial_{\mu}\varphi
\partial_{\nu}\varphi
-g_{\mu\nu}\left
\lbrack\frac{1}{2}g^{\rho\sigma}\partial_{\rho}\varphi\partial_{\sigma}
\varphi-\frac{m^2c^2}{2\hbar^2}\varphi^2-V(\varphi)\right \rbrack
\label{kge3}
\end{equation}
is the energy-momentum tensor of the SF with  $T_0^0=\epsilon$ the energy
density. The SF may represent the axion. The instanton
potential of the axion
\cite{pq,wittenV,vv} is
\begin{equation}
\label{kge4}
V(\varphi)=\frac{m^2cf^2}{\hbar^3}\left\lbrack 1-\cos\left
(\frac{\hbar^{1/2}c^{1/2}\varphi}{f}\right
)\right\rbrack-\frac{m^2c^2}{2\hbar^2}\varphi^2,
\end{equation}
where $m$ is the
mass of the axion and $f$ is the axion decay
constant. For this potential, the KG equation (\ref{kge1}) takes the form
\begin{equation}
\Box\varphi+\frac{m^2c^{3/2}f}{\hbar^{5/2}}\sin\left
(\frac{\hbar^{1/2}c^{1/2}\varphi}{f}\right
)=0.
\label{kge5}
\end{equation}
This is the general relativistic sine-Gordon equation. Considering
the
dilute limit $\varphi\ll f/\sqrt{\hbar c}$ (which is valid in
particular in the nonrelativistic limit  $c\rightarrow
+\infty$ considered below)\footnote{According to Eq. (\ref{kge30}), the axion
decay
constant $f$ scales as $c^{3/2}$.} and expanding
the cosine term in Eq. (\ref{kge4}) in Taylor series, we obtain at leading
order
the
$\varphi^4$ potential
\begin{equation}
\label{kge6}
V(\varphi)=-\frac{m^2c^3}{24
f^2\hbar}\varphi^4.
\end{equation}
In that case, the KG equation  (\ref{kge1}) takes the form 
\begin{equation}
\Box\varphi+\frac{m^2c^2}{\hbar^2}\varphi-\frac{m^2c^3}{6f^2\hbar}\varphi^3=0.
\label{kge7}
\end{equation}
In general, a quartic potential is written as
\begin{equation}
V(\varphi)=\frac{\lambda}{4\hbar c}\varphi^4,
\label{kge8}
\end{equation}
where $\lambda$ is the dimensionless self-interaction constant. Comparing Eqs.
(\ref{kge6}) and (\ref{kge8}), we find that
\begin{equation}
\lambda=-\frac{m^2c^4}{6f^2}.
\label{kge9}
\end{equation}
We note that $\lambda<0$, so that the $\varphi^4$
self-interaction term for axions is attractive. It leads to the collapse of
dilute axion stars above a maximum mass \cite{prd1,bectcoll}. The next order
$\varphi^6$ term 
has been considered in \cite{phi6} and turns out to be repulsive. This
repulsion, that occurs at high densities, may stop the collapse of dilute axion
stars and lead to the formation of dense axion stars \cite{braaten}.

In the weak-field gravity limit of general relativity
$\Phi/c^2\ll 1$, using the simplest form of the Newtonian gauge, the
Friedmann-Lema\^itre-Robertson-Walker (FLRW) line element is given
by
\begin{equation}
ds^2=c^2\left(1+2\frac{\Phi}{c^2}\right)dt^2-a(t)^2\left(1-2\frac{\Phi}{c^2}
\right)\delta_{ij}dx^idx^j,
\label{conf1}
\end{equation}
where $\Phi({\bf r},t)$ is the Newtonian potential and $a(t)$ is the scale
factor. In the Newtonian limit $\Phi/c^2\rightarrow 0$, the KGE equations
(\ref{kge1}) and
(\ref{kge2}) for the
inhomogeneous SF reduce to
\begin{equation}
\frac{1}{c^2}\frac{\partial^2\varphi}{\partial
t^2}+\frac{3H}{c^2}\frac{\partial\varphi}{\partial
t}-\frac{1}{a^2}\Delta\varphi+\frac{m^2c^2}{ \hbar^2 } \left
(1+\frac{2\Phi}{c^2}\right
)\varphi+\frac{dV}{d\varphi}=0,
\label{kge10}
\end{equation}
\begin{equation}
\frac{\Delta\Phi}{4\pi Ga^2}=\frac{\epsilon}{c^2}-\frac{3H^2}{8\pi G},
\label{kge11}
\end{equation}
where $H=\dot a/a$ is the Hubble parameter and the energy density is
given by
\begin{equation}
\epsilon=\frac{1}{2c^2}\left (\frac{\partial\varphi}{\partial t}\right
)^2+\frac{1}{2a^2}(\nabla\varphi)^2+\frac{m^2c^2}{2\hbar^2}\varphi^2+V(\varphi).
\label{kge12}
\end{equation}
For the $\varphi^4$ potential (\ref{kge6}), we get
\begin{equation}
\frac{1}{c^2}\frac{\partial^2\varphi}{\partial
t^2}+\frac{3H}{c^2}\frac{\partial\varphi}{\partial
t}-\frac{1}{a^2}\Delta\varphi+\frac{m^2c^2}{\hbar^2}\left
(1+\frac{2\Phi}{c^2}\right
)\varphi-\frac{m^2c^3}{6f^2\hbar}\varphi^3=0
\label{kge13}
\end{equation}
and
\begin{equation}
\epsilon=\frac{1}{2c^2}\left (\frac{\partial\varphi}{\partial t}\right
)^2+\frac{1}{2a^2}(\nabla\varphi)^2+\frac{m^2c^2}{2\hbar^2}\varphi^2-\frac{
m^2c^3 } {
24
f^2\hbar}\varphi^4.
\label{kge14}
\end{equation}
In equations (\ref{kge10})-(\ref{kge12}), we have neglected general relativity
(i.e. we have treated
gravity in the Newtonian framework)\footnote{This is valid
provided the
system is sufficiently far from forming a black hole.} but we have
kept special relativity
effects (see Eqs. (2) and (3) of \cite{phi6} for more
general equations
valid at the order $O(\Phi/c^2)$ in the post-Newtonian
approximation).
Considering now the nonrelativistic limit
$c\rightarrow +\infty$ where the SF displays rapid oscillations, these
equations can be simplified. To that
purpose, we write 
\begin{equation}
\label{kge15}
\varphi=\frac{1}{\sqrt{2}}\frac{\hbar}{m}\left\lbrack \psi({\bf
r},t)e^{-imc^2t/\hbar}+\psi^*({\bf r},t)e^{imc^2t/\hbar}\right\rbrack,
\end{equation}
where the complex wave function $\psi({\bf r},t)$ is a
slowly varying function of time (the fast oscillations $e^{imc^2t/\hbar}$ of the
SF have been factored out). This transformation allows us to separate the fast
oscillations of the SF with pulsation $\omega=mc^2/\hbar$
caused by its rest
mass from the slow evolution of $\psi({\bf r},t)$. From Eq. (\ref{kge15}), we
get
\begin{equation}
\dot\varphi=\frac{1}{\sqrt{2}}\frac{\hbar}{m}\left\lbrack
\dot\psi\, e^{-imc^2t/\hbar}-\frac{imc^2}{\hbar}
\psi\, e^{-imc^2t/\hbar}+{\rm c.c.}\right\rbrack,
\label{kge16}
\end{equation}
\begin{equation}
\nabla\varphi=\frac{1}{\sqrt{2}}\frac{\hbar}{m}\left (
\nabla\psi\, e^{-imc^2t/\hbar}+{\rm c.c.}\right ),
\label{kge17}
\end{equation}
\begin{eqnarray}
\ddot\varphi=\frac{1}{\sqrt{2}}\frac{\hbar}{m}\Biggl\lbrack
\ddot\psi\, e^{-imc^2t/\hbar}-\frac{2imc^2}{\hbar}
\dot\psi\, e^{-imc^2t/\hbar}\nonumber\\
-\frac{m^2c^4}{\hbar^2}
\psi\, e^{-imc^2t/\hbar}+{\rm c.c.}\Biggr\rbrack,
\label{kge18}
\end{eqnarray}
\begin{equation}
\Delta\varphi=\frac{1}{\sqrt{2}}\frac{\hbar}{m}\left (
\Delta\psi\, e^{-imc^2t/\hbar}+{\rm c.c.}\right ),
\label{kge19}
\end{equation}
where c.c.  denotes complex conjugaison. These equations are
exact.\footnote{For a noninteracting SF ($V=0$), substituting
Eqs. (\ref{kge15})-(\ref{kge19}) into Eq. (\ref{kge10}), we get the
exact special relativistic wave equation
\begin{equation}
\frac{1}{c^2}\frac{\partial^2\psi}{\partial
t^2}+\frac{3H}{c^2}\left
(\frac{\partial\psi}{\partial
t}-\frac{imc^2}{\hbar}\psi\right )-\frac{2im}{\hbar}\frac{\partial\psi}{\partial
t}-\frac{1}{a^2}\Delta\psi+\frac{2m^2}{\hbar^2}\Phi\psi=0.
\label{spe}
\end{equation}
} On the other
hand, if we compute ${\dot\varphi}^2$, $(\nabla\varphi)^2$, $\varphi^2$,
$\varphi^3$ and $\varphi^4$ from Eqs. (\ref{kge15})-(\ref{kge17})
and neglect terms with a rapidly oscillating phase factor
$e^{inmc^2t/\hbar}$ with $n\ge
2$, we get\footnote{Since we are considering the slowly varying
part of the wave function, we can remove all parts that oscillate with a
frequency much larger than $mc^2/\hbar$, i.e., we can neglect all parts that
change with a frequency $2mc^2/\hbar$, $3mc^2/\hbar$, $4mc^2/\hbar$... To a good
approximation,
we can argue
that the fast oscillating parts average to zero in the evolution of
$\varphi$. This eliminates particle number changing, as
discussed at the end of this Appendix.}
\begin{equation}
{\dot\varphi}^2=\frac{\hbar^2}{m^2}\left |\frac{\partial\psi}{\partial
t}\right |^2 +c^4|\psi|^2-2\frac{\hbar
c^2}{m}{\rm Im}\left (\frac{\partial\psi}{\partial
t}\psi^*\right ),
\label{kge22a1}
\end{equation}
\begin{equation}
(\nabla\varphi)^2=\frac{\hbar^2}{m^2}|\nabla\psi|^2,
\label{kge22a2}
\end{equation}
\begin{equation}
\varphi^2=\frac{\hbar^2}{m^2}|\psi|^2,
\label{kge22}
\end{equation}
\begin{equation}
\label{kge20}
\varphi^3\simeq \frac{1}{2\sqrt{2}}\frac{\hbar^3}{m^3}\left (
3\psi^2\psi^*  e^{-imc^2t/\hbar}+{\rm c.c.} \right ),
\end{equation}
\begin{equation}
\label{kge20phi4}
\varphi^4=\frac{3\hbar^4}{2m^4}|\psi|^4.
\end{equation}
Substituting these relations into the KGE equations
(\ref{kge11}), (\ref{kge13}) and (\ref{kge14}), and  neglecting oscillatory
terms, we obtain
the relativistic
GPP equations (see Eqs. (7)
and (8) of \cite{phi6} for more general
equations valid at the order $O(\Phi/c^2)$ in the post-Newtonian
approximation):
\begin{eqnarray}
i\hbar\frac{\partial\psi}{\partial
t}-\frac{\hbar^2}{2mc^2}\frac{\partial^2\psi}{\partial
t^2}-\frac{3}{2}H\frac{\hbar^2}{mc^2}\frac{\partial\psi}{\partial
t}+\frac{\hbar^2} { 2m a^2}
\Delta\psi\nonumber\\
-m\Phi\psi-m\frac{dV_{\rm
eff}}{d|\psi|^2}\psi+\frac{3}{2}i\hbar H\psi=0, 
\label{kge21}
\end{eqnarray}
\begin{eqnarray}
\frac{\Delta\Phi}{4\pi
Ga^2}=|\psi|^2+\frac{\hbar^2}{2m^2c^4}\left |\frac{\partial\psi}{
\partial t}\right |^2+\frac { \hbar^2 } { 2a^2m^2c^2 }
|\nabla\psi|^2\nonumber\\
+\frac{1}{c^2}V_{\rm
eff}(|\psi|^2)-\frac{\hbar}{mc^2}{\rm Im}\left (\frac{\partial\psi}{\partial
t}\psi^*\right )-\frac{3H^2}{8\pi G}\quad 
\label{gpoiss}
\end{eqnarray}
with the effective potential 
\begin{equation}
\label{kge27}
V_{\rm eff}(|\psi|^2)=-\frac{\hbar^3
c^3}{16f^2m^2}|\psi|^4.
\end{equation}
We note that $V_{\rm eff}(|\psi|^2)\equiv \overline{V(\varphi)}$ is different
from the expression
that one would have obtained by directly substituting Eq. (\ref{kge22}) into Eq.
(\ref{kge6}).\footnote{This is because
$\varphi$ is a real SF. Therefore, substituting $\varphi$ (exact) from Eq.
(\ref{kge15}) into $V(\varphi)$ {\it then} averaging over the oscillations is
different from substituting $\varphi$  (already averaged
over the oscillations) from Eq. (\ref{kge22}) into $V(\varphi)$. In other
words, $\overline{V_1(\varphi^2)}\neq V_1(\overline{\varphi^2})$ where we have
set $V(\varphi)=V_1(\varphi^2)$. The case of a complex SF $\varphi$ has been
considered in \cite{abrilph,playa}.
In that case, the transformation of the equations from $\varphi$ to $\psi$ is
exact (there is no need to average over the oscillations) and the potential is
unchanged ($V_{\rm eff}=V$).} They differ by a factor $2/3$. On the other
hand, assuming
$(\hbar/mc^2)|\ddot\psi|\ll |\dot\psi|$, 
$(\hbar/mc^2)|\dot\psi|\ll |\psi|$ and $mc^2/\hbar\gg H$, Eqs. (\ref{kge21})
and (\ref{gpoiss}) reduce to
\begin{equation}
i\hbar\frac{\partial\psi}{\partial
t}+\frac{3}{2}i\hbar H\psi=-\frac{\hbar^2}{2ma^2}
\Delta\psi+m\Phi\psi+m\frac{dV_{\rm
eff}}{d|\psi|^2}\psi, 
\label{rkge21}
\end{equation}
\begin{equation}
\frac{\Delta\Phi}{4\pi
Ga^2}=|\psi|^2+\frac { \hbar^2 } { 2a^2m^2c^2 }
|\nabla\psi|^2+\frac{1}{c^2}V_{\rm
eff}(|\psi|^2)-\frac{3H^2}{8\pi G}.
\label{rgpoiss}
\end{equation}
Finally, taking the nonrelativistic limit $c\rightarrow +\infty$, we obtain the
GPP
equations 
\begin{equation}
i\hbar\frac{\partial\psi}{\partial
t}+\frac{3}{2}i\hbar H\psi=-\frac{\hbar^2}{2ma^2}
\Delta\psi+m\Phi\psi+m\frac{dV_{\rm
eff}}{d|\psi|^2}\psi,
\label{rrkge21}
\end{equation}
\begin{equation}
\frac{\Delta\Phi}{4\pi
Ga^2}=|\psi|^2-\frac{3H^2}{8\pi G}.
\label{rrgpoiss}
\end{equation}
In the last expression, the energy density is given by
\begin{equation}
\frac{\epsilon}{c^2}\simeq |\psi|^2.
\label{kge23}
\end{equation}
Since $\epsilon/c^2$ represents, in the nonrelativistic
limit $c\rightarrow +\infty$, the rest-mass density
$\rho$, we conclude that
$\rho=|\psi|^2$. Therefore, the field equation (\ref{kge11}) reduces to
the
Poisson equation
\begin{equation}
\Delta\Phi=4\pi G a^2\left (\rho-\frac{3H^2}{8\pi G}\right ).
\label{kge24}
\end{equation}

Equation (\ref{kge27}) is
valid for the $\varphi^4$ potential (\ref{kge6}). More generally, the
effective
potential associated with the axion potential (\ref{kge4}) is
\begin{equation}
\label{kge28}
V_{\rm eff}(|\psi|^2)=\frac{m^2cf^2}{\hbar^3}\left\lbrack
1-\frac{\hbar^3 c}{2f^2m^2}|\psi|^2-J_0\left
(\sqrt{\frac{2\hbar^{3}c|\psi|^2}{f^2m^2}}\right
)\right\rbrack,
\end{equation}
where $J_0$ is the Bessel function of zeroth order (see
\cite{ebycollapse,phi6} for a detailed derivation). In
the dilute limit $|\psi|^2\ll f^2m^2/\hbar^3c$ (which is valid in particular in
the nonrelativistic limit $c\rightarrow +\infty$, see footnote
19) the
effective potential
$V_{\rm eff}(|\psi|^2)$ is dominated by the $|\psi|^4$ term. If we expand Eq.
(\ref{kge28})
in powers of $|\psi|$, we recover  Eq. (\ref{kge27}) at leading order. A
$|\psi|^4$ effective potential is
usually written as
\begin{equation}
\label{kge29}
V_{\rm eff}(|\psi|^2)=\frac{2\pi
a_s\hbar^2}{m^3}|\psi|^4,
\end{equation}
where $a_s$ is the s-scattering length of the bosons
\cite{revuebec}.\footnote{The ordinary GP equation with a cubic
nonlinearity \cite{gross1,gross2,gross3,pitaevskii2} is usually derived from the
mean field
Schr\"odinger equation \cite{bogoliubov} with a pair contact
potential \cite{hy,lhy} (see, e.g., Sec. II.A. of \cite{prd1}). The present
approach shows that the GP equation with a cubic nonlinearity may also
be derived from the KG equation with a quartic self-interaction potential.
More generally, the GP equation with a nonlinearity $V_{\rm
eff}(|\psi|^2)$ may
be derived from the KG equation with a self-interaction potential
$V(\varphi)$ (see Refs.  \cite{abrilph,playa}
and \cite{phi6} for a more detailed discussion of these issues in the case of a 
complex
or a real potential respectively).}
Comparing Eqs.
(\ref{kge27}) and (\ref{kge29}), we find that
\begin{equation}
\label{kge30}
a_s=-\frac{\hbar c^3m}{32\pi f^2}.
\end{equation}
We note that the scattering length is negative ($a_s<0$) corresponding to
an attractive self-interaction. On
the other hand,
comparing Eqs.
(\ref{kge9}) and (\ref{kge30}) yields\footnote{We note that the
relation between $\lambda$
and $a_s$ is different for a real SF and for a complex SF (see Appendix A of
\cite{bectcoll} for a complex SF). They differ by a factor $2/3$ for the reason
indicated in footnote 23.}
\begin{eqnarray}
\frac{\lambda}{8\pi}=\frac{2a_s m c}{3\hbar}.
\label{kge31}
\end{eqnarray}

The nonrelativistic limit $c\rightarrow +\infty$ can also be performed directly
in the action of the SF. Let us consider the nongravitational case for
brevity of presentation (we also assume a static background). In that case, the
action of the SF is 
\begin{equation}
S=\int d^4x\, {\cal L}
\label{lag1}
\end{equation}
with the Lagrangian density 
\begin{equation}
{\cal L}=
\frac{1}{2c^2}\left
(\frac{\partial\varphi}{\partial
t}\right
)^2-\frac{1}{2}(\nabla\varphi)^2-\frac{m^2c^2}{2\hbar^2}\varphi^2-V(\varphi).
\label{lag2}
\end{equation}
The least
action principle $\delta S=0$ leads to the Euler-Lagrange equation
\begin{equation}
\partial_{\mu}\left\lbrack \frac{\partial {\cal
L}}{\partial(\partial_{\mu}\varphi)}\right\rbrack-\frac{\partial{\cal
L}}{\partial\varphi}=0,
\label{lag3}
\end{equation}
yielding the KG equation 
\begin{equation}
\frac{1}{c^2}\frac{\partial^2\varphi}{\partial
t^2}-\Delta\varphi+\frac{m^2c^2}{ \hbar^2 }\varphi+\frac{dV}{d\varphi}=0.
\label{lag4}
\end{equation}
On the other hand, substituting Eqs. (\ref{kge22a1})-(\ref{kge20phi4}) into the
Lagrangian (\ref{lag2}) and neglecting oscillatory terms, we obtain the action
\begin{equation}
S=\int  {\cal L}\, d{\bf r} 
\label{lag5}
\end{equation}
with the Lagrangian
\begin{equation}
{\cal L}=\frac{\hbar^2}{2m^2c^2}\left |\frac{\partial\psi}{\partial t}\right
|^2-\frac{\hbar}{m}\, {\rm Im}\left (\frac{\partial\psi}{\partial
t}\psi^*\right )-\frac{\hbar^2}{2m^2}|\nabla\psi|^2-V_{\rm eff}(|\psi|^2).
\label{lag6}
\end{equation}
In the nonrelativistic limit $c\rightarrow +\infty$, it reduces to
\begin{equation}
{\cal L}=
\frac{\hbar}{2m} i \left
(\frac{\partial\psi}{\partial
t}\psi^*-\psi \frac{\partial\psi^*}{\partial
t}\right
)-\frac{\hbar^2}{2m^2}|\nabla\psi|^2-V_{\rm eff}(|\psi|^2),
\label{lag5b}
\end{equation}
where we have used
\begin{equation}
2i\, {\rm Im}\left (\frac{\partial\psi}{\partial
t}\psi^*\right )=\frac{\partial\psi}{\partial
t}\psi^*-\psi \frac{\partial\psi^*}{\partial
t}.
\label{lag6b}
\end{equation}
We recover the Lagrangian of a nonrelativistic BEC (see, e.g.,
Appendix B of \cite{bectcoll}). The Euler-Lagrange equation
\begin{equation}
\frac{\partial}{\partial t}\left (\frac{\partial{\cal
L}}{\partial\dot\psi}\right )+\nabla\cdot \left (\frac{\partial{\cal
L}}{\partial\nabla\psi}\right )-\frac{\partial{\cal L}}{\partial\psi}=0
\label{lag7}
\end{equation}
yields the GP equation 
\begin{equation}
i\hbar\frac{\partial\psi}{\partial
t}=-\frac{\hbar^2}{2m}
\Delta\psi+m\frac{dV_{\rm
eff}}{d|\psi|^2}\psi.
\label{lag8}
\end{equation}

{\it Remark:} Basically, axions are described by a relativistic quantum field
theory with a real scalar field $\varphi$ that obeys the KGE equations. In that
case, the particle number is not conserved. However, axions whose kinetic
energies are much
smaller than $mc^2$ can be described by a nonrelativistic effective field
theory with a complex SF $\psi$ that obeys the GPP equations. In that case,
they are just spinless particles whose number $N=\frac{1}{m}\int |\psi|^2\,
d{\bf r}$ is conserved. Physically, the particle number is conserved
because, by removing the fast oscillating terms, we have eliminated the particle
number violating processes that are energetically forbidden for
nonrelativistic particles.

\section{Ginzburg-Landau-Poisson, Cahn-Hilliard-Poisson and Smoluchowski-Poisson
equations}
\label{sec_gl}

In this Appendix, we consider Ginzburg-Landau-Poisson (GLP),
Cahn-Hilliard-Poisson (CHP) and Smoluchowski-Poisson (SP)  equations that can
serve as numerical algorithms to
compute stable equilibrium states of the GPP
equations.\footnote{Note that similar numerical algorithms,
having the form of generalized Fokker-Planck equations, have been introduced in
\cite{gen,nfp,assisi,vpre} in order to compute stable equilibrium states of the
Vlasov-Poisson
and Euler-Poisson equations.}

\subsection{Equations for $\psi$}

The GP equation can be written as \cite{chavtotal}
\begin{equation}
\label{gl1}
i\hbar\frac{\partial\psi}{\partial
t}=-\frac{\hbar^2}{2m}\Delta\psi+m\left\lbrack V'(|\psi|^2)+\Phi+\Phi_{\rm
ext}\right\rbrack \psi,
\end{equation}
where, for the sake of generality, we have considered an arbitrary potential of
self-interaction $V(|\psi|^2)$ and we have added an external potential
$\Phi_{\rm ext}({\bf r})$. We also recall that $\Phi({\bf r},t)$ is the
gravitational potential determined by the Poisson equation (\ref{gpp2}). More
generally, it can represent the mean field potential $\Phi({\bf r},t)=\int
u(|{\bf r}-{\bf r}'|)\rho({\bf r}',t)\, d{\bf r}'$ associated with a long-range
binary potential of interaction $u(|{\bf r}-{\bf r}'|)$. The energy
functional associated with the GPP equations is \cite{chavtotal}
\begin{eqnarray}
\label{gl2}
E_{\rm tot}=\frac{\hbar^2}{2m^2}\int |\nabla\psi|^2\, d{\bf r}+\int
V(|\psi|^2)\, d{\bf r}\nonumber\\
+\frac{1}{2}\int |\psi|^2 \Phi \, d{\bf r}+\int |\psi|^2 \Phi_{\rm
ext} \, d{\bf r}.
\end{eqnarray}
We have
\begin{equation}
\label{gl5}
i\hbar\frac{\partial\psi}{\partial
t}=m\frac{\delta E_{\rm tot}}{\delta\psi^*}.
\end{equation}
The GPP equations conserve the mass $M=\int |\psi|^2  \, d{\bf r}$ and
the energy $E_{\rm
tot}$. A stationary solution is obtained by extremizing $E_{\rm
tot}$ at fixed $M$, writing
$\delta E_{\rm tot}-\frac{\mu}{m}\delta M=0$. Since
\begin{eqnarray}
\label{gl3}
\frac{\delta E_{\rm
tot}}{\delta\psi^*}=-\frac{\hbar^2}{2m^2}\Delta\psi+\left\lbrack
V'(|\psi|^2)+\Phi+\Phi_{\rm
ext}\right\rbrack \psi,
\end{eqnarray}
we get
\begin{equation}
\label{gl4}
-\frac{\hbar^2}{2m}\Delta\psi+m\left\lbrack V'(|\psi|^2)+\Phi+\Phi_{\rm
ext}\right\rbrack \psi=\mu\psi.
\end{equation}
The same equation, with $\mu=E$ (eigenenergy), can be obtained by substituting 
$\psi({\bf r},t)=\phi({\bf r})
e^{-iEt/\hbar}$ into Eq. (\ref{gl1}). 
It can be shown that an equilibrium state is stable if, and only if, it is a
minimum of $E_{\rm tot}$ at
fixed $M$. In order to compute the stable steady states of the GP equation,
Huepe {\it et al.} \cite{huepe} propose to solve the GL equation\footnote{This
amounts to integrating the Schr\"odinger equation in imaginary time
\cite{gs1,gs2}.} 
\begin{equation}
\label{gl6}
-\hbar\frac{\partial\psi}{\partial
t}=m\frac{\delta F}{\delta\psi^*},
\end{equation}
where $F=E_{\rm tot}-\frac{\mu}{m} M$ is a grand potential. This equation can
be written explicitly as
\begin{equation}
\label{gl7}
-\hbar\frac{\partial\psi}{\partial
t}=-\frac{\hbar^2}{2m}\Delta\psi+m\left\lbrack V'(|\psi|^2)+\Phi+\Phi_{\rm
ext}\right\rbrack \psi-\mu\psi.
\end{equation}
The GLP equations satisfy an
H-theorem for the grand potential: $\dot F=-(2m/\hbar)\int |\delta
F/\delta\psi|^2\, d{\bf r}\le 0$. As a result,
they relax towards a steady state of the form of Eq. (\ref{gl4}) which
minimizes $F$ at fixed $\mu$. This is therefore a stable steady state of the GPP
equations with this value of $\mu$. The GL equation does
not conserve the mass $M$ (contrary to the GP equation). This is because it is
associated to a
grand canonical description where the chemical potential is fixed instead of the
mass. In order to obtain, at equilibrium, the correct value of $\mu$
corresponding to a prescribed mass $M$, Huepe {\it et
al.} \cite{huepe}  propose
to solve Eq. (\ref{gl7}) with a chemical potential $\mu(t)$ that evolves in time
so as to conserve $M$. This amounts to introducing formally a canonical
description
where the mass is fixed.

\subsection{Equations for $\rho$}

In the hydrodynamic representation, the energy
functional associated with the GPP equations is \cite{chavtotal}
\begin{eqnarray}
\label{gl8}
E_{\rm tot}=\frac{1}{m}\int \rho Q\, d{\bf r}+\int
V(\rho)\, d{\bf r}\nonumber\\
+\frac{1}{2}\int \rho \Phi \, d{\bf r}+\int \rho \Phi_{\rm
ext} \, d{\bf r},
\end{eqnarray}
where we have not written the classical kinetic term $\Theta_c=(1/2)\int \rho
{\bf u}^2\, d{\bf r}$ since we will be interested by equilibrium states only. 
The quantum hydrodynamic equations equivalent to the GPP equations can be
written in terms of functional derivatives of $E_{\rm tot}$ (see Sec. 3.6 of
\cite{chavtotal}). The GPP equations, or the corresponding hydrodynamic
equations, conserve
the mass
$M=\int \rho \, d{\bf r}$ and the energy $E_{\rm
tot}$ (including $\Theta_c$). A steady
state is obtained by extremizing $E_{\rm tot}$ at fixed $M$, writing
$\delta E_{\rm tot}-\frac{\mu}{m}\delta M=0$. Since
\begin{eqnarray}
\label{gl9}
\frac{\delta E_{\rm
tot}}{\delta\rho}=\frac{Q}{m}+V'(\rho)+\Phi+\Phi_{\rm ext},
\end{eqnarray}
we get
\begin{equation}
\label{gl10}
Q+m(V'(\rho)+\Phi+\Phi_{\rm ext})=\mu. 
\end{equation}
The same equation can be obtained from the condition of quantum hydrostatic
equilibrium (\ref{eq1}) using $V''(\rho)=h'(\rho)=P'(\rho)/\rho$, where $h$
is the enthalpy \cite{chavtotal}.  An equilibrium state is stable if, and only
if, it is a
minimum of $E_{\rm tot}$ at
fixed $M$.  In order
to compute a stable steady state of the GPP equations, we can solve the GL
equation 
\begin{equation}
\label{gl11}
\xi\frac{\partial\rho}{\partial
t}=-\frac{\delta F}{\delta\rho}
\end{equation}
or, explicitly,
\begin{equation}
\label{gl12}
-m\xi\frac{\partial\rho}{\partial
t}=Q+m(V'(\rho)+\Phi+\Phi_{\rm ext})-\mu,
\end{equation}
with the same comments as those following Eq. (\ref{gl7}).

{\it Remark:} An alternative manner to compute stable equilibrium states of the
GP equation is to solve the CH equation
\begin{equation}
\label{gl6b}
\xi\frac{\partial\rho}{\partial
t}=\Delta\frac{\delta E_{\rm tot}}{\delta\rho},
\end{equation}
which conserves mass and satisfies an
H-theorem for the energy: $\dot E_{\rm tot}=-(1/\xi)\int \lbrack\nabla (\delta
E_{\rm tot}/\delta\rho)\rbrack^2\, d{\bf r}\le 0$. Following \cite{chavtotal},
we may also consider the generalized CH equation
\begin{equation}
\label{gl6c}
\xi\frac{\partial\rho}{\partial
t}=\nabla\cdot \left ( \rho \nabla \frac{\delta E_{\rm
tot}}{\delta\rho}\right ),
\end{equation}
which conserves mass and satisfies an
H-theorem for the energy: $\dot E_{\rm tot}=-(1/\xi)\int \rho \lbrack\nabla
(\delta
E_{\rm tot}/\delta\rho)\rbrack^2\, d{\bf r}\le 0$. Explicitly, this equation
has the form of a quantum Smoluchowski equation \cite{chavtotal}
\begin{equation}
\label{gl6d}
\xi\frac{\partial\rho}{\partial
t}=\nabla\cdot \left (\nabla P+\rho\nabla\Phi+\rho\nabla\Phi_{\rm
ext}+\frac{\rho}{m}\nabla Q\right ).
\end{equation}
It corresponds to the strong friction limit $\xi\rightarrow +\infty$ of the
damped GP equation introduced in \cite{chavtotal}
\begin{eqnarray}
\label{gl6e}
i\hbar\frac{\partial\psi}{\partial
t}=-\frac{\hbar^2}{2m}\Delta\psi+m\left\lbrack V'(|\psi|^2)+\Phi+\Phi_{\rm
ext}\right\rbrack \psi\nonumber\\
-i\frac{\hbar}{2}\xi\left\lbrack
\ln\left (\frac{\psi}{\psi^*}\right
)-\left\langle \ln\left (\frac{\psi}{\psi^*}\right
)\right\rangle\right\rbrack\psi.
\end{eqnarray}
Therefore, this dissipative equation may serve as a numerical algorithm. 
The damped quantum Euler equations, equivalent to Eq.
(\ref{gl6e}), could be considered as well \cite{chavtotal}. In these different
exemples, the mass is automatically conserved so it is not necessary to enforce
its conservation with a Lagrange multiplier $\mu(t)$ as for the GL equation.

\subsection{Equations for $\phi$}

Let us set $\rho=\phi^2$ where $\phi$ is real. In that case, the energy
functional (\ref{gl8}) can be rewritten as
\begin{eqnarray}
\label{gl13}
E_{\rm tot}=\frac{\hbar^2}{2m^2}\int (\nabla\phi)^2\, d{\bf r}+\int
V(\phi^2)\, d{\bf r}\nonumber\\
+\frac{1}{2}\int \phi^2 \Phi \, d{\bf r}+\int \phi^2 \Phi_{\rm
ext} \, d{\bf r},
\end{eqnarray}
where, as before, we have not written the classical kinetic term $\Theta_c$. The
GPP equations conserve the mass $M=\int \phi^2 \, d{\bf r}$
 and the energy $E_{\rm
tot}$ (including $\Theta_c$). A steady
state is obtained by extremizing $E_{\rm tot}$ at fixed $M$, writing
$\delta E_{\rm tot}-\frac{\mu}{m}\delta M=0$. Since
\begin{eqnarray}
\label{gl14}
\frac{\delta E_{\rm
tot}}{\delta\phi}=-\frac{\hbar^2}{m^2}
\Delta\phi+2V'(\phi^2)\phi+2\phi\Phi+2\phi\Phi_ { \rm ext},
\end{eqnarray}
we get
\begin{equation}
\label{gl15}
-\frac{\hbar^2}{2m}
\Delta\phi+m\left\lbrack V'(\phi^2)+\Phi+\Phi_ { \rm
ext}\right\rbrack \phi=\mu\phi.
\end{equation}
An
equilibrium state is stable if, and only if, it is a
minimum of $E_{\rm tot}$ at
fixed $M$.  In order
to compute a stable steady state of the GPP equations, we can solve the GL
equation 
\begin{equation}
\label{gl16}
\xi\frac{\partial\phi}{\partial
t}=-\frac{\delta F}{\delta\phi}
\end{equation}
or, explicitly,
\begin{equation}
\label{gl17}
-\frac{\xi m}{2}\frac{\partial\phi}{\partial
t}=-\frac{\hbar^2}{2m}\Delta\phi+m\left\lbrack V'(\phi^2)+\Phi+\Phi_ { \rm
ext}\right\rbrack \phi-\mu\phi,
\end{equation}
with the same comments as those following Eq. (\ref{gl7}).

\section{Thermal tunneling 
in the stochastic Ginzburg-Landau equation}
\label{sec_sgl}

In this Appendix, we take thermal fluctuations into account in the
framework of the stochastic GL
equation \cite{goldenfeld}.\footnote{This equation has been
studied
numerically recently by Verma {\it et al.} \cite{verma} in relation to
self-gravitating BECs.} We compute the thermal tunneling rate of a field
$\rho({\bf r},t)$ across a barrier
of free energy by using the instanton theory \cite{bray}. Our approach
provides, in this context,
a justification of the Kramers formula giving the typical lifetime of a
metastable state. Similar results can be obtained for the stochastic CH and
generalized CH (or Smoluchowski)
equations \cite{cdrandom,entropy,entropy2}.

The stochastic GL equation writes 
\begin{eqnarray}
\label{glch1}
\xi\frac{\partial\rho}{\partial t}=-\frac{\delta F}{\delta\rho}+\sqrt{2\xi
k_{B}T}\zeta({\bf r},t),
\end{eqnarray}
where $\zeta({\bf r},t)$ is a Gaussian white noise. The free energy
$F[\rho]$  can be an arbitrary functional of $\rho$, but it is usually written
under
the form 
\begin{eqnarray}
\label{glch2}
F\lbrack{\rho}\rbrack=\int \left\lbrack \frac{1}{2}
(\nabla{\rho})^{2}+V({\rho})\right\rbrack d{\bf r}.
\end{eqnarray}
The potential $V(\rho)$ is also an arbitrary function of $\rho$ but it is often
approximated by its normal form close to a critical point according to the
Landau theory of
phase transitions.  For a
functional of the form of Eq. (\ref{glch2}),  the 
stochastic GL equation (\ref{glch1}) can be
written explicitly as
\begin{equation}
\label{glch3}
\xi\frac{\partial\rho}{\partial t}=\Delta\rho-V'(\rho)+\sqrt{2\xi
k_{B}T}\zeta({\bf r},t).
\end{equation}
In the absence of noise ($T=0$),  the deterministic GL equation writes
\begin{eqnarray}
\label{glch1b}
\xi\frac{\partial\rho}{\partial t}=-\frac{\delta
F}{\delta\rho}=\Delta\rho-V'(\rho).
\end{eqnarray}
Its  equilibrium
states  are extrema of $F$:
\begin{eqnarray}
\label{glch3c}
\frac{\delta F}{\delta\rho}=0\qquad \Leftrightarrow \qquad
-\Delta\rho+V'(\rho)=0.
\end{eqnarray}
On the other hand, it satisfies an H-theorem
\begin{eqnarray}
\label{glch3b}
\dot F=\int \frac{\delta F}{\delta\rho}\frac{\partial\rho}{\partial t}\, d{\bf
r}=-\frac{1}{\xi}\int  \left (\frac{\delta F}{\delta\rho}\right )^2\, d{\bf
r}\le 0.
\end{eqnarray}
As a result, the deterministic GL equation relaxes towards a stable equilibrium
state which minimizes $F$
(maxima or saddle points are linearly unstable).
In the presence of noise ($T\neq 0$), the stochastic GL
equation (\ref{glch1}) can be interpreted as a Langevin equation. 
The probability density
$P[\rho,t]$ of the density field $\rho({\bf r},t)$ at time $t$ is governed by
the functional FP equation
\begin{equation}
\label{glch4}
\xi\frac{\partial P}{\partial t}[\rho,t]
=\int d{\bf r}\, \frac{\delta}{\delta\rho({\bf r})}\left\lbrace\left\lbrack k_B
T\frac{\delta}{\delta\rho({\bf r})}+\frac{\delta F}{\delta\rho({\bf
r})}\right\rbrack
P[\rho,t]\right\rbrace.
\end{equation}
It relaxes  towards the equilibrium Boltzmann distribution
\begin{eqnarray}
P[\rho]=\frac{1}{Z(\beta)}e^{-\beta F[\rho]}.
\label{glch4b}
\end{eqnarray}

We assume that the free
energy functional $F[\rho]$ has a local minimum  $\rho_M({\bf
r})$ (metastable state) and a global minimum $\rho_S({\bf
r})$
(stable state) separated by a maximum or a saddle point $\rho_U({\bf
r})$ (unstable state). In the absence
of noise,
the evolution of the system is deterministic and the density relaxes towards one
of
the minima of
the potential  as implied by the $H$-theorem (\ref{glch3b}). In the
presence of noise, the
density switches back and forth between the two
minima (attractors). When the  noise is
weak ($T\rightarrow 0$),\footnote{For systems with long-range interactions, the
noise is also weak when $N\rightarrow +\infty$.}
the
transition between the two
minima is a rare event.
One important problem is to determine the rate $\Gamma$ for the density profile,
initially located in the metastable state $\rho_M({\bf
r})$, to cross the barrier of free energy 
and
reach the stable state $\rho_S({\bf
r})$.

Since the distribution of the Gaussian white noise $\zeta({\bf r},t)$ is
\begin{eqnarray}
P[\zeta({\bf r},t)]\propto e^{-\int_{-\infty}^{+\infty}dt\int d{\bf r}\,
\zeta^2({\bf r},t)/2},
\label{in1}
\end{eqnarray}
the probability of the path $\rho({\bf r},t)$ is
\begin{eqnarray}
P[\rho({\bf r},t)]\propto e^{-S[\rho({\bf r},t)]/k_B T},
\label{in1b}
\end{eqnarray}
where $S$ is the generalized Onsager-Machlup (OM) functional \cite{om}
\begin{eqnarray}
S[\rho({\bf r},t)]=\frac{1}{4\chi}\int dt\int d{\bf r}
\, \left (\frac{\partial\rho}{\partial t}+\chi \frac{\delta
F}{\delta\rho}\right )^2.
\label{in2}
\end{eqnarray}
The functional $S$ may be called an action by analogy with the
path-integral
formulation of quantum mechanics (the temperature $T$ plays the role of the
Planck constant $\hbar$ in quantum mechanics) \cite{feynman}.  It can be written
as $S=\int L\, dt$ where $L$ is the corresponding Lagrangian.
The probability density to observe the system with the profile $\rho_2({\bf
r})$
at time $t_2$ given that it had the profile $\rho_1({\bf
r})$ at time $t_1$ is
\begin{eqnarray}
\label{in3}
P[\rho_2({\bf r}),t_2|\rho_1({\bf r}),t_1]= \int {\cal D}\rho\, 
e^{-S[\rho]/k_B T},
\end{eqnarray}
where the integral runs over all paths satisfying $\rho({\bf r},t_1)=\rho_1({\bf
r})$ and $\rho({\bf r},t_2)=\rho_2({\bf
r})$. For a given initial condition $\rho_0({\bf r})$ at $t=t_0$, the
probability density $P[\rho({\bf r}),t]\equiv P[\rho({\bf r}),t|\rho_0({\bf
r}),t_0]$ to observe the system with the profile $\rho({\bf
r})$
at time $t$ satisfies the functional FP equation (\ref{glch4}).
In the
weak noise
limit, the typical paths explored by the
system are concentrated close to the most probable path. In that case, a
steepest-descent evaluation of the path integral is possible. The path
integral is dominated by the most probable path. To
determine the most probable path, we have to minimize the
OM functional $S[\rho({\bf r},t)]$, i.e., we have to solve the
minimization problem
\begin{eqnarray}
\label{ommin1}
\min_{\rho({\bf r},t)}\quad \lbrace S[\rho({\bf r},t)] \rbrace.
\end{eqnarray}
The equation for the most
probable path $\rho_c({\bf r},t)$ that connects two attractors is called an
``instanton'' \cite{instanton}. It is obtained by cancelling the first order
variations of the
action
\begin{eqnarray}
\label{ommin2}
\delta S=0.
\end{eqnarray}
In the weak
noise
limit, the transition probability from one state to the other
is dominated by the most probable path:
\begin{eqnarray}
\label{in4}
P[\rho_2({\bf r}),t_2|\rho_1({\bf r}),t_1]\simeq e^{-S[\rho_c]/k_B T}.
\end{eqnarray}
This formula can be interpreted as a large deviation result. It provides an
approximate solution of the functional FP equation (\ref{glch4}). On
the other hand, it can be shown that the
escape
rate of the system
over the barrier of free energy is given by
\begin{eqnarray}
\label{in5}
\Gamma \propto e^{-S[\rho_c]/k_B T},
\end{eqnarray}
where $S[\rho_c]$ is the action of the most probable path (instanton) that
connects 
the metastable state to the stable state. In the limit of weak noise,
it can be shown that the most
probable
path between the metastable state and the stable state must necessarily pass
through the saddle point  $\rho_U({\bf
r})$  (playing the role of a ``critical droplet'' in
problems of nucleation).
Once
the system reaches the saddle point it may either return to the initial
metastable state or reach the stable state. In the latter
case, it has
crossed the barrier of free energy.

To determine the instanton which solves the variational problem (\ref{ommin1}),
we can
proceed as follows. The Lagrangian associated with the OM functional (\ref{in2})
is
\begin{eqnarray}
L=\frac{1}{4\chi}\int d{\bf r}
\, \left (\frac{\partial\rho}{\partial t}+\chi \frac{\delta
F}{\delta\rho}\right )^2.
\label{in6}
\end{eqnarray}
The corresponding Hamiltonian is defined by 
\begin{eqnarray}
H=\int \dot\rho \frac{\delta L}{\delta\dot\rho}\, d{\bf r}-L.
\label{in7}
\end{eqnarray}
Since the Lagrangian does not explicitly depend on time, the Hamiltonian is
conserved. Therefore, using Eq. (\ref{in6}), we get
\begin{eqnarray}
H=\frac{1}{4\chi}\int d{\bf r}
\, \left (\frac{\partial\rho}{\partial t}-\chi \frac{\delta
F}{\delta\rho}\right )\left (\frac{\partial\rho}{\partial t}+\chi \frac{\delta
F}{\delta\rho}\right ),
\label{in8}
\end{eqnarray}
where $H$ is a constant.
Since the attractors satisfy $\partial\rho/\partial t=0$ and $\delta
F/\delta\rho=0$, the constant $H$ is equal to zero ($H=0$). Therefore,
the instanton satisfies the equations
\begin{eqnarray}
\label{in9}
\frac{\partial\rho_c}{\partial t}=\mp
\chi \frac{\delta F}{\delta\rho_c}
\end{eqnarray}
with the boundary conditions $\rho_c({\bf r},-\infty)=\rho_M({\bf
r})$ and $\rho_c({\bf r},+\infty)=\rho_S({\bf r})$.
We note that the most probable path corresponds to the deterministic dynamics
(\ref{glch1b}) with a sign $\mp$.\footnote{Considering
the
solution with the sign $-$, which corresponds to the downhill solution (see
below), we see that the most probable path (instanton) coincides with
the ensemble average path, i.e., the deterministic GL equation
(\ref{glch1b}) obtained by averaging the stochastic GL equation
(\ref{glch1}) over the noise. It has a zero action ($S=0$). As a
result, the deterministic GL
equation (\ref{glch1b}) -- the average path -- can be obtained by
minimizing the OM functional
(\ref{in2}).} The physical interpretation of Eq.
(\ref{in6}) is the following. Starting from the metastable state, the most
probable path follows the time-reversed deterministic dynamics against the free
energy
gradient up to the saddle point; beyond the saddle point, it follows the
forward-time deterministic dynamics down to the stable state. According to Eqs.
(\ref{in2}) and (\ref{in9}), the action of the most
probable path corresponding to the transition from the saddle point to the
stable state (downhill solution corresponding to Eq. (\ref{in9}) with the
sign $-$) is zero  while the action of the most probable path  corresponding to
the transition from the metastable state to the saddle point (uphill solution
corresponding to Eq. (\ref{in9}) with the sign $+$) is nonzero. This is to be
expected since the descent from the saddle point to the stable state is a
``free'' descent that does not require thermal noise; it thus gives the
smallest possible value of zero of the action. By contrast, the rise from the
metastable state to the saddle point is a rare event that requires thermal
noise. The action for the
uphill solution is
\begin{equation}
S[\rho_c({\bf r},t)]=\int dt\int d{\bf r} \, \frac{\partial\rho_c}{\partial
t}\frac{\delta
F}{\delta\rho_c}
=\int dt\, \frac{dF}{dt}=\Delta F,
\label{in10}
\end{equation}
where $\Delta F=F[\rho_U]-F[\rho_M]$ is the
barrier of free energy between the
metastable state and the unstable state.
The total action for the most
probable path connecting the attractors is therefore
$S_c=S[\rho_c^{-}]+S[\rho_c^{+}]=\Delta F+0=\Delta F$. It is determined solely
by the
uphill path. The instanton solution gives the dominant contribution to the
transition rate for a weak noise. Therefore, the rate for the system to
pass
from the metastable state to the stable state (escape rate) is
\begin{eqnarray}
\label{in11}
\Gamma\propto e^{-\Delta F/k_B T}.
\end{eqnarray}
This is the celebrated Arrhenius (or Kramers) formula stating that the
transition rate
is inversely proportional to the exponential of the barrier of free energy
divided by $k_B T$.\footnote{This formula can be
simply
obtained as follows. The equilibrium probability of observing the density
$\rho({\bf
r})$ is
$\propto e^{-\beta F[\rho]}$. Therefore, the probability for the system
initially prepared in the metastable state to form a ``critical droplet''
(unstable state $\rho_U$) and then reach the stable state $\rho_S$ is $\propto
e^{-\beta (F[\rho_U]-F[\rho_M])}$. The typical lifetime of a metastable state
may then be estimated by $t_{\rm life}\sim e^{\beta\Delta F}$, where $\Delta
F=F[\rho_U]-F[\rho_M]$ is the
barrier of free energy between the
metastable state and the unstable state.}
The
typical lifetime of a metastable state is $t_{\rm
life}\sim\Gamma^{-1}$. For systems with long-range
interactions, the free
energy scales as $N$ so the typical lifetime of a metastable state scales as 
\begin{eqnarray}
t_{\rm life}\propto e^{N \Delta f/k_B T}.
\end{eqnarray}
For systems with long-range interactions, the metastable states are
very relevant since their lifetime scales as $e^N$ with $N\gg 1$. Therefore,
metastable states are stable in practice.
Only very close to the critical point where $\Delta f\rightarrow 0$ does their
lifetime decrease substantially.

\section{Maximum mass of general relativistic self-interacting boson stars}
\label{sec_mgrdas}

We consider a  relativistic complex SF $\varphi$ with a self-interaction
potential $V(|\varphi|^2)$ like in Refs.
\cite{colpi,chavharko,mlbec,abrilph,playa,shapiro,abrilphas,shapironew,guerra}.
In the TF (or semiclassical) limit where the quantum kinetic energy can
be neglected, the energy density and the pressure are given by \cite{abrilphas}
\begin{eqnarray}
\epsilon=\rho c^2+V(\rho)+\rho V'(\rho),
\label{mgrdas1}
\end{eqnarray}
\begin{eqnarray}
P=\rho V'(\rho)-V(\rho),
\label{mgrdas2}
\end{eqnarray}
where $\rho$ is the pseudo rest-mass density
\begin{eqnarray}
\rho=\frac{m^2}{\hbar^2}|\varphi|^2.
\label{mgrdas3}
\end{eqnarray}
Therefore, in this approximation, a
self-interacting boson star is equivalent to a relativistic fluid described by a
barotropic
equation of state $P(\epsilon)$ defined in implicit form by Eqs. (\ref{mgrdas1})
and (\ref{mgrdas2}). We note that Eq. (\ref{mgrdas2}) has the same form as in
the nonrelativistic limit where $\rho=|\psi|^2$ represents the mass density (see
\cite{chavtotal} for detail).

Let us consider a power-law potential 
\begin{eqnarray}
V(|\varphi|^2)=A|\varphi|^{2\gamma}
\label{mgrdas4}
\end{eqnarray}
with $\gamma>1$. Using Eq. (\ref{mgrdas3}), we get
\begin{eqnarray}
V(\rho)=\frac{K}{\gamma-1}\rho^{\gamma}
\label{mgrdas5}
\end{eqnarray}
with 
\begin{eqnarray}
K=(\gamma-1)A\left (\frac{\hbar}{m}\right )^{2\gamma}.
\label{mgrdas6}
\end{eqnarray}
According to Eq. (\ref{mgrdas2}), the pressure is given by
\begin{eqnarray}
P=K\rho^{\gamma}.
\label{mgrdas7}
\end{eqnarray}
This is a polytropic equation of state with polytropic constant $K$ and
polytropic index $\gamma=1+1/n$. On the other hand, according to Eq.
(\ref{mgrdas1}), the energy density is given
by 
\begin{eqnarray}
\epsilon=\rho c^2+\frac{K(\gamma+1)}{\gamma-1}\rho^{\gamma}=\rho
c^2+(2n+1)P.
\label{mgrdas8}
\end{eqnarray}
At low densities
$\rho\rightarrow 0$, we get $\epsilon\sim \rho c^2$ so that the energy density
is dominated by the rest-mass energy. This corresponds to the nonrelativistic
limit. At high densities $\rho\rightarrow +\infty$, we obtain $\epsilon\sim
(2n+1)P$ or, equivalently,
\begin{eqnarray}
P\sim \frac{1}{2n+1}\epsilon.
\label{mgrdas9}
\end{eqnarray}
This corresponds to the ultrarelativistic limit. Since the relation between the
pressure and the energy density is linear ($P=q\epsilon$), the
mass-radius relation $M(R)$, parametrized by $\epsilon$, forms a spiral at
high densities as in the case
of neutron stars \cite{htww}. Furthermore, the series of equilibria becomes
unstable at the maximum mass $M_{\rm max}$ correponding to the first turning
point of the spiral.\footnote{More precisely, a mode of stability is lost at a
turning point of mass if the $M(R)$ curve rotates anticlockwise (and
gained if it rotates clockwise) \cite{htww}. On the other hand, we know that
nonrelativistic polytropic gaseous spheres are stable for $n<3$ and unstable for
$n>3$ \cite{chandrabook}. Therefore, when $n<3$, the series of equilibria is
stable before the first turning point of mass and becomes unstable afterwards.
When $n>3$, the whole series of equilibria is unstable.} The square of the speed
of sound is
$c_s^2=P'(\epsilon)c^2=c^2/(2n+1)$. Since $n>0$, the speed  of
sound is always less than the speed of light ($c_s<c$).

(i) We first consider a $|\varphi|^4$ potential with a repulsive
self-interaction ($a_s>0$) of the form
\cite{colpi,chavharko,mlbec,abrilph,playa,shapiro,abrilphas,shapironew}
\begin{eqnarray}
V(|\varphi|^2)=\frac{2\pi a_s m}{\hbar^2}|\varphi|^{4}.
\label{mgrdas10}
\end{eqnarray}
Using Eq. (\ref{mgrdas3}) we get
\begin{eqnarray}
V(\rho)=\frac{2\pi a_s\hbar^2}{m^3}\rho^2.
\label{mgrdas11}
\end{eqnarray}
The pressure is given by
\begin{eqnarray}
P=\frac{2\pi a_s\hbar^2}{m^3}\rho^2.
\label{mgrdas12}
\end{eqnarray}
This is a polytropic equation of state of polytropic constant $K={2\pi
a_s\hbar^2}/{m^3}$ and polytropic index $\gamma=2$ (i.e. $n=1$). The energy
density is given by
\begin{eqnarray}
\epsilon=\rho
c^2+3P=\rho
c^2+\frac{6\pi a_s\hbar^2}{m^3}\rho^2
\label{mgrdas13}
\end{eqnarray}
This is quadratic equation for $\rho$. Solving this equation and substituting
the result into Eq. (\ref{mgrdas12}), we obtain the equation of state
\begin{eqnarray}
P=\frac{m^3c^4}{72\pi a_s\hbar^2}\left (\sqrt{1+\frac{24\pi
a_s\hbar^2}{m^3c^4}\epsilon}-1\right )^2.
\label{mgrdas14}
\end{eqnarray}
It coincides with the result of \cite{colpi}. For $\rho\rightarrow +\infty$,
the equation of state reduces to $P\sim \epsilon/3$ like for the
ordinary radiation (due to photons). The mass-radius
relation corresponding to the equation of state (\ref{mgrdas13}) has been
obtained in \cite{chavharko,mlbec}. It displays a maximum mass
\begin{equation}
\label{mgrdas14b}
M_{\rm max, GR}=0.307 \, \frac{\hbar c^2 \sqrt{a_s}}{(G m)^{3/2}}
\end{equation}
at a radius
\begin{equation}
\label{mgrdas14c}
R_{\rm *, GR}=1.923\left
(\frac{a_s\hbar^2}{Gm^3}\right )^{1/2}.
\end{equation}
and forms a spiral at high densities as explained previously.

(ii) We now consider axion boson stars (in the sense of \cite{guerra}) with the
axion boson potential $V(|\varphi|^2)$ truncated at the order $|\varphi|^6$ as
in \cite{phi6}:
\begin{eqnarray}
V(|\varphi|^2)=\frac{2\pi a_s m}{\hbar^2}|\varphi|^4+\frac{32\pi^2
a_s^2}{9\hbar^2c^2}|\varphi|^6.
\label{mgrdas15}
\end{eqnarray}
The $|\varphi|^4$ term is attractive ($a_s<0$) while the $|\varphi|^6$ is
repulsive. We are interested in describing the branch of dense axion boson
stars for large mass $M$ where general relativistic
effects are important. Since we are considering a complex SF, the number of
bosons is conserved. As a result, dense axion boson stars
should be stable with respect to the decay via emission of relativistic
axions contrary to the case where the SF is real (see the
introduction). Using Eq. (\ref{mgrdas3}) we get
\begin{eqnarray}
V(\rho)=\frac{2\pi a_s\hbar^2}{m^3}\rho^2+\frac{32\pi^2
a_s^2\hbar^4}{9m^6c^2}\rho^3.
\label{mgrdas16}
\end{eqnarray}
At high densities the repulsive 
$|\varphi|^6$ term dominates over the attractive $|\varphi|^4$. If we just
keep the repulsive $|\varphi|^6$ potential, we obtain
\begin{eqnarray}
V(\rho)=\frac{32\pi^2
a_s^2\hbar^4}{9m^6c^2}\rho^3.
\label{mgrdas17}
\end{eqnarray}
The pressure is given by
\begin{eqnarray}
P=\frac{64\pi^2
a_s^2\hbar^4}{9m^6c^2}\rho^3
\label{mgrdas18}
\end{eqnarray}
This is the equation of state of a polytrope with polytropic constant
$K={64\pi^2a_s^2\hbar^4}/{9m^6c^2}$ and polytropic index $\gamma=3$ (i.e.
$n=1/2$). The energy density is given by
\begin{eqnarray}
\epsilon=\rho c^2+\frac{128\pi^2
a_s^2\hbar^4}{9m^6c^2}\rho^3=\rho
c^2+2P.
\label{mgrdas19}
\end{eqnarray}
This is a third degree equation for $\rho$.  For $\rho\rightarrow +\infty$,
the equation of state reduces to $P\sim \epsilon/2$. The corresponding
mass-radius relation will be studied in a specific paper \cite{prep}. We
just provide below preliminary results.

At low densities, the system is nonrelativistic.  The general
mass-radius relation of polytropic spheres is
\begin{equation}
\label{mgrdas20}
M^{(n-1)/n}R^{(3-n)/n}=\frac{K(1+n)}{(4\pi)^{1/n}G}\omega_n^{(n-1)/n}, 
\end{equation}
where $\omega_n$ is a constant that can be obtained from the Lane-Emden
equation \cite{chandrabook}.
Specializing on the equation of state (\ref{mgrdas18}), we obtain
\begin{equation}
\label{mgrdas21}
M=\frac{3Gm^6c^2}{2\hbar^4a_s^2}\omega_{1/2}R^5=0.0323 \frac{Gm^6c^2}{
\hbar^4a_s^2}R^5,
\end{equation}
where we have used $\omega_{1/2}=0.02156...$. We note that the mass increases
with the radius.

At high densities, the system is ultrarelativistic. Since the equation of state
is
linear at high densities, we expect that the mass-radius relation $M(R)$ will
form a spiral and display a  maximum mass $M_{\rm max}$.  An estimate
of the maximum mass of general relativistic
dense boson axion stars in the $|\varphi|^6$ approximation can be obtained by
combining the Newtonian mass-radius
relation
(\ref{mgrdas21}) with the constraint $R\ge R_S$, where $R_S=2GM/c^2$ is the
Schwarzschild
radius. This
gives a maximum general relativistic
mass 
\begin{equation}
\label{mgrdas22}
M_{\rm max, GR}^{\rm dense}=0.991 \, \left
(\frac{|a_s|\hbar^2c^4}{G^3m^3}\right )^{1/2}
\end{equation}
and a correponding radius
\begin{equation}
\label{mgrdas23}
R_{\rm *,GR}^{\rm dense}=1.98\left
(\frac{|a_s|\hbar^2}{Gm^3}\right )^{1/2}.
\end{equation}
We can also express these results in
terms of the axion decay constant
\begin{equation}
\label{mgrdas24}
f=\left (\frac{\hbar c^3m}{32\pi |a_s|}\right )^{1/2}.
\end{equation}
We get
\begin{equation}
\label{mgrdas25}
M_{\rm max, GR}^{\rm dense}=0.0988 \, \left
(\frac{\hbar^3c^7}{G^3}\right )^{1/2}\frac{1}{f m},
\end{equation}
\begin{equation}
\label{mgrdas26}
R_{\rm *, GR}^{\rm dense}= 0.197 \, \left
(\frac{\hbar^3c^3}{G}\right )^{1/2}\frac{1}{f m}.
\end{equation}
If we measure the axion decay constant $f$ in units of $10^{15}{\rm GeV}$ and 
 the axion mass $m$ in units of $10^{-22}\, {\rm eV/c^2}$ we get
$M_{\rm max, GR}^{\rm dense}=1.61\times 10^{15}\, (f m)^{-1}\,
M_{\odot}$ and  $R_{\rm *, GR}^{\rm dense}=154 \, (f m)^{-1}\,
{\rm pc}$.

For QCD axions with  $m=10^{-4}\,
{\rm eV}/c^2$, $a_s=-5.8\times 10^{-53}\, {\rm m}$ and $f=5.82\times
10^{19}\, {\rm eV}=4.77\times 10^{-9} M_P c^2$,  we obtain
 $M_{\rm max, GR}^{\rm dense}=27.7\, M_{\odot}$ and $R_{\rm *,GR}^{\rm
dense}=81.9\, {\rm
km}$.

For ULAs with $m=2.19\times 10^{-22}\, {\rm eV}/c^2$, $a_s=-1.11\times
10^{-62}\, {\rm fm}$ and $f=1.97\times 10^{23}\, {\rm eV}=1.61\times 10^{-5} M_P
c^2$, we obtain $M_{\rm max, GR}^{\rm
dense}=3.74\times 10^{15}\, M_{\odot}$ and $R_{\rm *,GR}^{\rm dense}=358\, {\rm
pc}$.

{\it Remark:} For QCD axions, the product $m f\equiv(\Lambda_{\rm QCD}/c)^2$ of
the mass
and decay constant is fixed to the value $\Lambda_{\rm QCD}=7.6\times 10^7\,
{\rm
eV}$ \cite{kc}.  
This gives a universal maximum mass and maximum stable radius $M_{\rm max,
GR}^{\rm dense}=27.7\, M_{\odot}$ and $R_{\rm *,GR}^{\rm
dense}=81.9\, {\rm
km}$. We stress that this result is valid only for the $|\varphi|^6$
potential given by Eq. (\ref{mgrdas17}) in the TF limit for which $M_{\rm max,
GR}^{\rm
dense}\propto 1/(f m)$. The fact that the maximum mass obtained numerically by
Guerra {\it et al.} \cite{guerra} depends on $f$ when $m f$ is fixed shows
that the rigorous description of dense axion boson stars is more complicated
than the
present analysis.

\end{document}